\documentclass[11pt]{article}
\usepackage{amsfonts,amsthm,amsmath,amssymb,latexsym,comment}

\usepackage{times}
\pdfoutput=1
\ifx\pdftexversion\undefined
\usepackage[colorlinks,linkcolor=black,filecolor=black,citecolor=black,urlcolor=black,pdfstartview=FitH]{hyperref}
\else
  \usepackage[colorlinks,linkcolor=blue,filecolor=blue,citecolor=blue,urlcolor=blue,pdfstartview=FitH]{hyperref}
\fi
\usepackage{epsfig}
\usepackage{graphicx}

\newcommand{\wbox}{\mbox{$\sqcap$\llap{$\sqcup$}}}

\newtheorem{theorem}{Theorem}

\newtheorem{lemma}{Lemma}
\newtheorem{proposition}[lemma]{Proposition}
\newtheorem{definition}{Definition}
\newtheorem{example}{Example}

\newcommand{\citeyear}{\cite}

\newcommand{\namedref}[2]{\hyperref[#2]{#1~\ref*{#2}}}
\newcommand{\sectionref}[1]{\namedref{Section}{#1}}
\newcommand{\appendixref}[1]{\namedref{Appendix}{#1}}

\newcommand{\theoremref}[1]{\namedref{Theorem}{#1}}

\newcommand{\figureref}[1]{\namedref{Figure}{#1}}

\newcommand{\lemmaref}[1]{\namedref{Lemma}{#1}}

\newcommand{\propref}[1]{\namedref{Proposition}{#1}}
\newcommand{\dagree}{{\em Detect/Agree\ }}

\renewcommand{\P}{\mathcal{P}}

\newcommand{\sigmact}{\sigma_{\mbox{\footnotesize\sc ct}}}
\newcommand{\Gammact}{\Gamma_{\!\mbox{\footnotesize\sc ct}}}
\newcommand{\vecsigmact}{\sigma_{\mbox{\footnotesize\sc ct}}}

\newcommand{\uT}{u^T}

\def\beginsmall#1{\vspace{-\parskip}\begin{#1}\itemsep-\parskip}
\def\endsmall#1{\end{#1}\vspace{-\parskip}}
\newcommand{\R}{\mathcal{R}}
\newenvironment{RETHM}[2]{\trivlist \item[\hskip 10pt\hskip\labelsep
\hskip -10pt{\bf
#1\hskip 5pt\relax\ref{#2}.}]\it}{\endtrivlist}

\newcommand{\rethm}[1]{\begin{RETHM}{Theorem}{#1}}
\newcommand{\repro}[1]{\begin{RETHM}{Proposition}{#1}}
\newcommand{\relem}[1]{\begin{RETHM}{Lemma}{#1}}
\newcommand{\recor}[1]{\begin{RETHM}{Corollary}{#1}}

\newcommand{\erethm}{\end{RETHM}}
\newcommand{\erepro}{\end{RETHM}}
\newcommand{\erelem}{\end{RETHM}}
\newcommand{\erecor}{\end{RETHM}}

\newcommand{\T}{{\cal T}}
\newcommand{\eps}{\mbox{$\epsilon$--}}

\newcommand{\fullv}[1]{\commentout{#1}}
\renewcommand{\fullv}[1]{#1}

\newcommand{\commentout}[1]{}

\newcommand{\punish}{{\textsc{punish}}}
\newcommand{\pass}{{\textsc{pass}}}
\newcommand{\detect}{{\textsc{detect}}}

\title{Lower Bounds on Implementing Robust and Resilient Mediators}
\author{
Ittai Abraham\\
School of Computer Science and Engineering\\
The Hebrew University of Jerusalem\\
Jerusalem, Israel\\
\texttt{ittaia@cs.huji.ac.il}
\and
Danny Dolev\thanks{Part of the work was done
while the author visited Cornell university. The work was
funded in part by ISF, ISOC, NSF, CCR, and AFOSR.}\\
School of Computer Science and Engineering\\
The Hebrew University of Jerusalem\\
Jerusalem, Israel\\
\texttt{dolev@cs.huji.ac.il}
\and
Joseph Y. Halpern\thanks{
Supported in part by NSF under grants CCR-0208535, ITR-0325453, and
IIS-0534064, by ONR under grant N00014-01-10-511, by the DoD
Multidisciplinary University Research Initiative (MURI) program
administered by the ONR under grants N00014-01-1-0795 and
N00014-04-1-0725, and by AFOSR under grant FA9550-05-1-0055.
} \\
Cornell University\\
Ithaca, NY 14850\\
\texttt{halpern@cs.cornell.edu}}

\begin{document}

\maketitle
\begin{abstract}
We provide new and tight lower bounds on the ability of players to
implement equilibria using cheap talk, that is, just allowing
communication among the players. One of our main results
is
that,
in general,
 it is impossible to implement
three-player Nash equilibria in a bounded number of rounds.
We also give the first rigorous connection between Byzantine agreement lower
bounds and lower bounds on implementation. To this end we consider a
number of variants of
Byzantine agreement
and introduce reduction
arguments. We also give lower bounds on the running time of two player
implementations. All our results extended to lower bounds on
\emph{$(k,t)$-robust} equilibria, a solution concept that tolerates
deviations by coalitions of size up to $k$ and deviations by up to $t$
players with unknown utilities
(who may be malicious).
\end{abstract}

\thispagestyle{empty}
\newpage
\setcounter{page}{1}

\section{Introduction}
The question of whether a problem in a multiagent system that can be
solved with a trusted mediator can be solved by just the agents in the
system, without the mediator, has attracted a great deal of attention in
both computer science (particularly in the cryptography community) and
game theory.  In cryptography, the focus on the problem has been on
\emph{secure multiparty computation}.  Here it is assumed that each
agent $i$ has some private information $x_i$.  Fix functions $f_1,
\ldots, f_n$.  The goal is have agent $i$ learn $f_i(x_1, \ldots, x_n)$
without learning anything about $x_j$ for $j \ne i$ beyond what is
revealed by the value of $f_i(x_1, \ldots, x_n)$.
With a trusted mediator, this is trivial: each agent $i$ just gives
the mediator its private value $x_i$; the mediator then sends each
agent $i$ the value $f_i(x_1, \ldots, x_n)$.  Work on multiparty
computation (see \cite{goldreich03} for a survey) provides
conditions under which this can be done.  In game theory, the focus
has been on whether an equilibrium in a game with a mediator can be
implemented using what is called \emph{cheap talk}---that is, just
by players communicating among themselves (see \cite{M97} for a
survey).

There is a great deal of overlap between the problems studied in
computer science and game theory.  But there are some significant
differences.  Perhaps the most significant difference is that, in
the computer science literature, the interest has been in doing
multiparty computation in the presence of possibly malicious
adversaries, who do everything they can to subvert the computation.
On the other hand, in the game theory literature, the assumption is
that players have preference and seek to maximize their utility;
thus, they will subvert the computation iff it is in their best
interests to do so.
Following \cite{ADGH06}, we consider here both rational adversaries, who
try to maximize their utility, and possibly malicious adversaries (who
can also be
considered
rational adversaries whose utilities we do not
understand).

\subsection{Our Results}\label{intro1}
In this paper we provide new and optimal lower bounds on the ability to
implement mediators with cheap talk. Recall
that a \emph{Nash equilibrium} $\sigma$ is a tuple of strategies such
that given that all other players play their corresponding part of
$\sigma$ then the best response is also to play $\sigma$. Given a Nash
equilibrium $\sigma$ we say that a strategy profile $\rho$ is a
\emph{$k$-punishment strategy for $\sigma$} if, when all
but $k$ players play their component of $\rho$, then no matter what the
remaining $k$ players do, their payoff is strictly less than what it is
with $\sigma$.
We now describe some highlights of our results in the two simplest
settings: (1) where rational players cannot form coalitions and there
are no malicious players (this gives us the solution concept of Nash
equilibrium) and (2) where there is at most one malicious player.
We describe our results in a more general setting in
\sectionref{intro2}.

\paragraph{No bounded implementations:}
In \cite{ADGH06} it was shown that any Nash equilibrium with a
mediator for three-player games with a 1-punishment strategy can be
implemented using cheap talk. The expected running time of
the
implementation is constant. It is natural to ask if implementations with
a bounded number
of
rounds exist for all three-player
games. \theoremref{thm:lowerbound-4} shows this is not the case,
implementations must have infinite executions and cannot be bounded for
all three-player games. This lower bound highlights the importance of
using randomization.
An earlier attempt to provide a three-player cheap talk implementation
\cite{Bp03} uses a bounded implementation, and hence cannot work in general.
The key insight of the lower bound is that when the implementation
is bounded, then at some point the punishment strategy
must become ineffective. The details turn out to be quite subtle.
The only other lower bound that we are aware of that has the same flavor is the
celebrated FLP result~\cite{FLP}
for reaching agreement in
asynchronous systems, which also shows that no bounded implementation
exists. However, we use quite different proof techniques than FLP.

\paragraph{Byzantine Agreement and Game Theory:}
We give the first rigorous connection between Byzantine agreement lower
bounds and lower bounds on
implementation.
To get the lower bounds, we need to consider a number of variants of
Byzantine agreement, some novel.
The novel variants require new impossibility results.
We have four results of this flavor:
\begin{enumerate}
\item  Barany~\cite{Barany92} gives an example to show that, in general,
to implement an equilibrium with a mediator in a three-player game,
it is necessary to have a 1-punishment strategy.
Using the power of randomized Byzantine
agreement lower bounds we strengthen his result and show
in \theoremref{thm:lowerbound-3} that
we cannot even get an $\epsilon$-implementation in this setting.

\item Using the techniques of \cite{BGW88} or \cite{F90}, it is easy to
show that any four-player
game Nash equilibrium with a mediator can be implemented using cheap
talk even if no 1-punishment strategy exists. Moreover, these
implementations are \emph{universal}; they do not depend on the players'
utilities. In \theoremref{thm:lowerbound-2} we prove that universal
implementations do not exist
in general for three-player games. Our proof uses a nontrivial reduction
to the weak Byzantine agreement (WBA) problem~\cite{L83}. To obtain our lower bound, we need
to prove a new impossibility
result
for WBA, namely,
that no protocol with a finite expected running time
can solve WBA.

\item In \cite{ADGH06} we show that for
six-player
games with a
2-punishment strategy, any Nash equilibrium can be implemented even in the
presence of at most one malicious player. In
\theoremref{thm:lowerboundb} we show that for five players even
$\eps$-implementation is impossible. The proof uses a
variant of Byzantine agreement; this  is related to the problem of
\emph{broadcast with
extended consistency} introduced by Fitzi~et al.~\citeyear{FHHW03}. Our
reduction maps the rational player to a Byzantine process that is
afraid of being detected and the malicious player to a standard
Byzantine process.

\item  In \theoremref{thm:d-broadcast-lowerbound}, we show that for
four-player games with at most one malicious player,
to implement the mediator, we must have a PKI setup in place,
even if the players are all computationally bounded and even
if we are willing to settle for $\eps$implementations. Our lower bound
is based on a reduction to
a novel relaxation of the Byzantine agreement
problem.
\end{enumerate}

\paragraph{Bounds on running time:}
We provide bounds on the number of rounds needed to implement
two-player games. In \theoremref{thm:d-lowerbound}(a) we
prove that the expected running time of any implementation of a
two-player mediator equilibrium must depend on the utilities of the
game, even if there is a 1-punishment
strategy. This is in contrast to the three-player case, where the
expected running time is constant. In \theoremref{thm:d-lowerbound}(b)
we prove that the expected running time of any $\eps$implementation of a
two-player mediator equilibrium
for which there is no 1-punishment strategy must
depend on $\epsilon$. Both results are obtained using a new
two-player variant of the secret-sharing game. The only result that we
are aware of that has a similar sprit is that of
Boneh and Naor~\cite{BN00}, where it is shown that two-party protocols with
``bounded unfairness'' of $\epsilon$ must have running time that depends
on the value of $\epsilon$.
The implementations given by Urbano and Vila \citeyear{UV02,UV04} in the
two-player case are independent of the utilities;
the above
results show
that
their implementation
cannot be correct in general.

\subsection{Our results for implementing robust and resistent
mediators}\label{intro2}

In \cite{ADGH06} (ADGH from now on), we argued that it is important to
consider deviations
by both rational players, who have preferences and try to maximize them,
and players
that can be viewed
as malicious, although it is perhaps better to
think of them as rational players whose utilities are
not known by the other players or mechanism designer.  We considered
equilibria
that are
\emph{$(k,t)$-robust}; roughly speaking, this means that the equilibrium
tolerates deviations by up to $k$ rational players, whose utilities are
presumed known,
and up to $t$ players with unknown utilities
(i.e., possibly malicious players).
We showed how $(k,t)$-robust equilibria with mediators could be
implemented using cheap talk, by first showing
that, under appropriate assumptions, we could implement secret
sharing in a $(k,t)$-robust way using cheap talk.
These assumptions involve standard considerations in the game
theory and distributed systems literature,
specifically,
(a) the relationship between $k$, $t$ and $n$, the total number
of players in the system; (b) whether players know the exact utilities
of other players; (c) whether there are broadcast channels  or just
point-to-point channels; (d) whether cryptography is available; and (e)
whether the game has a $(k+t)$-\emph{punishment strategy}; that is, a
strategy that, if used by all but at most $k+t$ players,
guarantees that every player gets a worse outcome than they do with the
equilibrium strategy.
Here we provide a complete picture of when implementation is
possible, providing lower bounds that match the known upper bounds
(or improvements of them that we have obtained).  The following is a
high-level picture of the results.
(The results discussed in \sectionref{intro1} are special cases of the
results stated below.  Note that all the upper bounds mentioned
here are either in ADGH, slight improvements of results in ADGH,
or
are known in the literature; see
\sectionref{sec:upperbound}
for the
details.  The new results
claimed in the current submission
are the matching lower bounds.)
\beginsmall{itemize}
\item If $on > 3k + 3t$, then mediators can be implemented using cheap
talk; no punishment strategy is required, no knowledge of other
agents' utilities is required, and the cheap-talk strategy has
bounded running time
that does not depend on the utilities
(\theoremref{thm:upperbound}(a)
in \sectionref{sec:upperbound}).
\item If $n \le 3k + 3t$, then we cannot, in general, implement a mediator
using cheap talk without  knowledge of other agents' utilities
(\theoremref{thm:lowerbound-2}).  Moreover, even if other agents'
utilities are
known, we cannot, in general, implement a mediator without having a
punishment strategy (\theoremref{thm:lowerbound-3}) nor with bounded
running time (\theoremref{thm:lowerbound-4}).

\item If $n > 2k+3t$, then mediators can be implemented using cheap talk
if there is a punishment strategy (and utilities are known) in
finite expected running time that does not depend on the utilities
(\theoremref{thm:upperbound}(b)
in \sectionref{sec:upperbound}).

\item If $n \le 2k+3t$, then we cannot, in general, $\epsilon$-implement a
mediator  using cheap talk, even if there is a punishment strategy
and utilities are known (\theoremref{thm:lowerboundb}).

\item If $n > 2k+2t$ and we can simulate broadcast then,
for all $\epsilon$,
we can
$\epsilon$-implement a mediator
using cheap talk, with bounded expected running time that does not
depend on the utilities in the game or
on $\epsilon$
(\theoremref{thm:upperbound}(c)
in \sectionref{sec:upperbound}).
(Intuitively, an $\epsilon$-implementation is an implementation
where a player can gain at most $\epsilon$ by deviating.)
\item If $n \le 2k+2t$, we cannot, in general, $\epsilon$-implement a
mediator using cheap talk even if we have broadcast channels
(\theoremref{pro:lowerboundbii}).
Moreover, even if we assume cryptography and broadcast channels, we
cannot, in general, $\epsilon$-implement a mediator using cheap talk
with expected running time that does not depend on $\epsilon$
(\theoremref{thm:d-lowerbound}(b)); even if there is a punishment
strategy, then we still cannot, in general, $\epsilon$-implement a
mediator using cheap talk with expected running time independent of
the utilities in the game (\theoremref{thm:d-lowerbound}(a)).

\item If $n > k+3t$ then, assuming cryptography, we can
$\epsilon$-implement a mediator using cheap talk;
moreover, if there is a punishment strategy, the expected running
time does not depend on $\epsilon$ (\theoremref{thm:upperbound}(e)
in \sectionref{sec:upperbound}).

\item
If $n \le k+3t$, then even assuming cryptography, we cannot, in
general, $\epsilon$-implement a mediator using cheap talk
(\theoremref{thm:d-broadcast-lowerbound}).

\item If $n > k+t$, then assuming cryptography and that
a PKI (Public Key Infrastructure) is in place,%
\footnote{We can replace the assumption of a PKI here and elsewhere by
the assumption that there is a trusted preprocessing phase where players
may broadcast.}
we can $\epsilon$-implement a
mediator (\theoremref{thm:upperbound}(d)
in \sectionref{sec:upperbound});
moreover, if there is a punishment strategy, the expected running
time does not depend on $\epsilon$ (\theoremref{thm:upperbound}(e)
in \sectionref{sec:upperbound}).

\endsmall{itemize}

The lower bounds are existential results; they show that if certain
conditions do not hold, then  there exists an
equilibrium that can be implemented by a mediator that cannot be
implemented using cheap talk.  There are other games where these
conditions do not hold but we can nevertheless implement a mediator.

\subsection{Related work}

There has been a great deal of work on implementing mediators,
both in computer science and game theory.  The results above generalize
a number of results that appear in the literature.  We briefly discuss
the most relevant work on implementing mediators here.  Other work
related to this paper is discussed where it is relevant.

In game theory, the study of implementing mediators using cheap talk goes
back to Crawford and Sobel \citeyear{CS82}.
Barany \citeyear{Barany92} shows that if $n \ge 4$, $k=1$, and $t=0$
(i.e., the setting for Nash equilibrium), a mediator
can be implemented in a game where players do not have private
information.  Forges \citeyear{F90}
provides what she calls a \emph{universal mechanism} for implementing
mediators; essentially, when combining her results with those of Barany,
we get the special case of \theoremref{thm:upperbound}(a) where $k=1$
and $t=0$.
Ben-Porath \cite{Bp03} considers implementing a mediator with cheap
talk %
in the case that $k=1$ if $n \ge 3$ and there is a 1-punishment
strategy. He seems to have been the first to consider punishment
strategies (although his notion is different from ours: he requires that
there be an equilibrium that is dominated by the equilibrium that we are
trying to implement).
Heller \cite{Heller05} extends Ben-Porath's result to allow
arbitrary $k$. \theoremref{thm:upperbound}(b) generalizes Ben-Porath
and Heller's results.  Although \theoremref{thm:upperbound}(b) shows
that the statement of Ben-Porath's result is correct, Ben-Porath's
implementation takes a bounded number of
rounds;  \theoremref{thm:lowerbound-4} shows it cannot be correct.%
\footnote{Although Heller's implementation does not take a bounded number
of rounds, it suffers from problems similar to those of Ben-Porath.}
Heller proves a matching lower bound; \theoremref{thm:lowerboundb}
generalizes Heller's lower bound to the case that $t > 0$.
(This turns out to require a much more complicated game than that
considered by Heller.)
Urbano
and Vila \cite{UV02,UV04} use cryptography to deal with the case that
$n=2$ and $k=1$;%
\footnote{However, they make somewhat vague and
nonstandard assumptions about the cryptographic tools they use.}
\theoremref{thm:upperbound}(e)) generalizes their
result to arbitrary $k$ and $t$.  However, just as with Ben-Porath,
Urbano and Vila's implementation takes a bounded number of rounds;
As we said in \sectionref{intro1},
\theoremref{thm:d-lowerbound}(a) shows that it cannot be correct.

In the cryptography community,
results on implementing mediators go
back to 1982 (although this terminology was not used), in the
context of \emph{(secure) multiparty computation}.  Since there are
no utilities in this problem, the focus has been on essentially what
we call here \emph{$t$-immunity}: no group of $t$ players can
prevent the remaining players from learning the function value, nor
can they learn the other players' private values. Results of Yao
\citeyear{yao:sc} can be viewed as showing that if $n=2$
and appropriate computational hardness assumptions are made,
then, for all $\epsilon$,  we can obtain $1$-immunity with probability
greater than $1-\epsilon$
if appropriate computational hardness
assumptions hold.  Goldreich, Micali, and Wigderson \cite{GMW87}
extend Yao's result to the case that $t > 0$ and $n > t$.
Ben-Or, Goldwasser, and Wigderson \citeyear{BGW88} and
Chaum, Cr{\' e}peau, and
Damgard \citeyear{CCD88} show that, without computational hardness
assumptions, we can get $t$-immunity if $n > 3t$; moreover,
the protocol of Ben-Or, Goldwasser, and Wigderson does not
need an $\epsilon$ ``error'' term.  Although they did not consider
utilities, their protocol actually gives a $(k,t)$-robust implementation
of a mediator using cheap talk if $n > 3k + 3t$; that is, they
essentially prove
\theoremref{thm:upperbound}(a).
(Thus, although these results predate those of Barany and Forges, they
are actually stronger.)
Rabin and Ben-Or \citeyear{RB89}  provide  a $t$-immune
implementation of a mediator with ``error'' $\epsilon$ if broadcast can
be simulated.  Again, when we add utilities, their protocol actually
gives an $\epsilon$--$(k,t)$-robust implementation.  Thus, they
essentially prove \theoremref{thm:upperbound}(c).
Dodis, Halevi, and Rabin \citeyear{DHR00} seem to have been the first
to apply cryptographic techniques to game-theoretic solution concepts; they
consider the case that
$n=2$ and $k=1$ and there is no private information (in which case the
equilibrium in the mediator game is a \emph{correlated equilibrium}
\cite{Aumann87}); their result is essentially that of Urbano and Vila
\citeyear{UV04} (although their protocol does not suffer form the
problems of that of Urbano and Vila).

Halpern and Teague \citeyear{HT04} were perhaps the first to
consider the general problem of multiparty computation with rational
players. In this setting, they essentially prove
\theoremref{thm:upperbound}(d) for the case that $t=0$ and $n \ge
3$. However, their focus is on the solution concept of
\emph{iterated deletion}.  They show that there is no Nash
equilibrium for rational multiparty computation with rational agents
that survives iterated deletion and give a protocol with finite
expected running time that does survive iterated deletion.
If $n \le 3(k+t)$, it follows easily from
\theoremref{thm:lowerbound-4}: that there is no multiparty
computation protocol that is a Nash equilibrium, we do not have to
require that the protocol survive iterated deletion to get the
result if $n \le  3(k+t)$.  Various generalizations of the Halpern
and Teague results have been proved.  We have already mentioned the
work of ADGH. Lysanskaya and Triandopoulos \citeyear{LT06}
independently proved the special case of
\theoremref{thm:upperbound}(c) where $k=1$ and $t+1<n/2$ (they also
consider survival of iterated deletion); Gordon and Katz
\citeyear{GK06} independently proved a special case of
\theoremref{thm:upperbound}(d)
where $k=1$, $t=0$, and $n\ge 2$.

In this paper we are interested in implementing equilibrium by using
standard communication channels. An alternate option is to consider
the possibility of simulating equilibrium by using much stronger
primitives.
Izmalkov, Micali, and Lepinski \citeyear{IML05} show that,
if there is a punishment strategy and we have available strong
primitives that they call \emph{envelopes} and \emph{ballot boxes},
we can implement arbitrary mediators perfectly (without an
$\epsilon$ error) in the case that $k=1$, in the sense that every
equilibrium of the game with the mediator corresponds to an
equilibrium of the cheap-talk game, and vice versa.
In \cite{LMPS04,LMS05}, these primitives are also used to
obtain implementation that is perfectly collusion proof in
the model where, in the game with the mediator, coalitions cannot
communicate.  (By way of contrast, we allow coalitions to communicate.)
Unfortunately,
envelopes and ballot boxes cannot be implemented under standard
computational and systems assumptions \cite{LMS05}.
\fullv{
It is reasonable to ask at this point whether mediators are of
practical interest.  After all, if three companies negotiate, they
can just hire an arguably trusted mediator, say an auditing firm.
The disadvantage of this approach
in a setting like the internet, with constantly shifting alliances,
there are always different groups that want to collaborate; a group
may not have the time and flexibility of hiring a mediator, even
assuming they can find one they trust.  Another concern is that our
results simply shift the role of what has to be trust elsewhere.  It
is certainly true that our results assume point-to-point
communication that cannot be intercepted. If
$n \le k + 3t$, then we must also assume the existence of a
public-key infrastructure. Thus, we have essentially shifted from
trusting the mediator to trusting the PKI.  In practice, individuals
who want to collaborate may find point-to-point communication and a
PKI more trustworthy than an intermediary, and easier to work with.
}

\commentout{
 For example, if $3k + 2t < n$,
$\sigma$ is a $(k,t)$-robust equilibrium strategy in a game with a
mediator, and there is a $(k+t)$-punishment strategy with respect to
$\sigma$, then we can implement $\sigma$ using cheap talk. ADGH
conjectured that there were lower bounds that matched our upper
bounds.  Here we essentially prove this result. For example, we show
that if $3k+2t \ge n$, then there exists a game with a
$(k,t)$-robust equilibrium that cannot be implemented using cheap
talk. In the process of proving some of the lower bounds, we were
able to improve the upper bounds.  Although our focus here is on
lower bounds, we explain the improved upper bounds here as well. }

The rest of this paper is organized as follows.  In
\sectionref{sec:definitions}, we review
the relevant
definitions.
In \sectionref{sec:upperbound}, we briefly discuss the upper bounds,
and compare them to the results of ADGH.
In \sectionref{sec:lowerbounds}, we prove the lower bounds.
\fullv{The missing proofs appear in the appendix.}

\section{Definitions}\label{sec:definitions}

In this section, we give detailed definitions of the main notions needed
for our results.  Sometimes there are subtle differences between our
differences and those used in the game-theory literature.  We discuss
these differences carefully.

\subsection{Mediators and cheap talk}

We are interested in implementing mediators.  Formally, this means
we need to consider three games: an \emph{underlying game} $\Gamma$,
an extension $\Gamma_d$ of $\Gamma$ with a mediator, and a
cheap-talk extension
$\Gammact$
of $\Gamma$.
Our underlying games are \emph{(normal-form) Bayesian games}.  These
are games of incomplete information, where players make only one
move, and these moves are made simultaneously. The ``incomplete
information''
is
captured
by assuming that nature makes the first move and
chooses for each player $i$ a \emph{type} in some set $\T_i$,
according to some distribution that is commonly known. Formally, a
Bayesian game $\Gamma$ is defined by a tuple $(N,\mathcal{T}, A, u,
\mu)$, where $N$ is the set of players,
$\T=\times_{i \in N} \T_i$ is the set of possible types,
$\mu$ is the distribution on types,  $A= \times_{i \in
N} A_i$ is the set of action profiles, and $u_i:\T \times A$ is the
utility of player $i$ as a function of the types prescribed by
nature  and the actions taken by all players.

A \emph{strategy} for player $i$ in a Bayesian game $\Gamma$ is a
function from $i$'s type to an action in $A_i$; in a game with a
mediator, a strategy is a function from $i$'s type and message
history to an action.  We allow behavior strategies (i.e.,
randomized strategies); such a strategy gets an extra argument,
which is a sequence of coin flips (intuitively, what a player does
can depend on its type, the messages it has sent and received if we
are considering games with mediators, and the outcome of some coin
flips).  We use lower-case Greek letters such as $\sigma$, $\tau$,
and $\rho$ to denote a strategy profile; $\sigma_i$ denotes the
strategy of player $i$ in strategy profile $\sigma$; if $K \subseteq
N$, then $\sigma_K$ denotes the strategies of  the players in $K$
and $\sigma_{-K}$ denotes the strategies of the players not in $K$.
Given a strategy profile $\sigma$ a player $i \in N$ and a type $t_i
\in T_i$ let $u_i(t_i,\sigma)$ be the expected utility of player $i$
given that his type is $t_i$ and each player $j \in N$ is playing
the strategy $\sigma_j$.

Given an underlying Bayesian game $\Gamma$ as above,
a game $\Gamma_d$ with a mediator $d$ that extends
$\Gamma$ is, informally, a game where players can communicate with
the mediator and then perform an action from $\Gamma$.
The utility function of a player $i$ in $\Gamma_d$ is the same as that
in $\Gamma$; thus, the utility of a player $i$ in $\Gamma_d$ depends
just on the types of all players and the actions taken by all players.
Formally, we view both the mediator and the players as interacting
Turing machines with access to an unbiased coin (which thus
allows them to choose uniformly at random from a finite set of any
size).  
The mediator and players interact 
for some (possibly unbounded) number of stages.  A mediator is
characterized by a function $\P$ that maps the inputs it has
received up to a stage (and possible some random bits) to an output
for each player.
Given an underlying Bayesian game $\Gamma$
where player $i$'s actions come from the set $A_i$ and a mediator $d$,
the interaction with the mediator in $\Gamma_d$
proceeds in stages, where each stage
consists of three phases. In the first phase of a stage, each player
$i$ sends an input to $d$ (player $i$ can send the empty input,
i.e., no input at all); in the second phase, $d$ sends each player
$i$ an output according to $\P$, (again, the mediator can send the
empty output); and in the third phase, each player $i$ chooses an
action in $A_i$ or no action at all. A player can
play at most one action from $A_i$ in each execution (play) of $\Gamma_d$.
Player $i$'s utility function in $\Gamma_d$ is the same as that in
the underlying game $\Gamma$, and depends only on the action profile
in $A$ played by the players and the types.  To make this precise,
we need to define what move an action in $A_i$ is played by player
$i$ in executions of $\Gamma_d$ where $i$ in fact never plays an
action in $A_i$.  For ease of exposition, we assume that for each
player $i$, some default action $a_i^* \in A$ is chosen.  There are
other ways of dealing with this issue (see, for example, \cite{AH03}
for an alternative approach).  Our results do not depend on the
choice, since in our upper bounds, with probability 1, all players
do play an action in equilibrium, and our impossibility results are
independent of the action chosen if players do not choose an action.
(We remark that the question of what happens after an infinite execution
of the cheap-talk game becomes much more significant in asynchronous
systems; see \cite{ADH07a}.)

Although we think of a \emph{cheap-talk} game as a game where players
can communicate with each other (using point-to-point communication and
possibly
broadcast), formally, it is a game with a special kind of
mediator:
player $i$ send the mediator whatever
messages it wants to send other players in the first phase of a
round; the mediator forwards these messages to the intended
recipients in the second phase.  We can model broadcast messages by
just having the mediator tag a message as a broadcast (and sending
the same message to all the intended recipients, of course).

We assume that cheap talk games are always unbounded; players are
allowed to talk forever.
The \emph{running time} of an execution of a joint strategy $\sigma$ is
the number of steps
taken until the last player makes a move in the underlying game.
The running time may be infinite.

It is standard in
the game theory literature to view cheap talk as \emph{pre-play}
communication.  Thus, although different plays of $\Gammact$ may
have different running times (possibly depending on random factors),
it is assumed that it is
commonly known when the cheap-talk phase ends; then all players make
their decisions simultaneously.  It is not possible for some players
to continue communicating after some other players have decided
(see, for example, \cite{Heller05}, where this assumption is
explicit). For
their possibility results, ADGH define games and give recommended
strategies for these games such that, as long as players use the
recommended strategy, all players make a move at the same time in
each play of the cheap-talk game.   However, it is not assumed that
this is the case off the equilibrium path (that is, if players do
not follow the recommended strategy).  The assumption that all
players stop communicating at the same time seems to us very strong,
and not implementable in practice,
so we do not make it here.  (Dropping this assumption sometimes
makes our impossibility results harder to prove; see
the proof of
\theoremref{pro:lowerboundbii} in
\appendixref{thm6proof} for an example.)  Thus, there is
essentially only one cheap-talk game extending an underlying game
$\Gamma$,;
$\Gammact$ denotes
the cheap-talk extension of $\Gamma$.

When we consider a deviation by a coalition $K$, we want to allow
the players in $K$ to communicate with each other.
If $\Gamma'$ is
an extension of an underlying game $\Gamma$ (including $\Gamma$
itself) and $K \subseteq N$, let $\Gamma'+CT(K)$ be the extension of
$\Gamma$ where the mediator provides private cheap-talk channels for
the players in $K$ in addition to whatever communication there is in
$\Gamma'$.  Note that $\Gammact + CT(K)$ is just $\Gammact$;
players in $K$ can already talk to each other in $\Gammact$.

\subsection{Implementation}

Note that a strategy profile---whether it
is in the underlying game, or in a game with a mediator extending
the underlying game (including a cheap-talk game)---induces a
mapping from type profiles to distributions over action profiles. If
$\Gamma_1$ and $\Gamma_2$ are extension of some underlying game
$\Gamma$, then strategy $\sigma_1$ in $\Gamma_1$ \emph{implements} a
strategy $\sigma_2$ in $\Gamma_2$ if both $\sigma$ and $\sigma'$
induce the same function from types to distributions over actions.
Note that although our informal discussion in the introduction talked
about \emph{implementing mediators}, the formal definitions (and our
theorems) talk about implementing strategies.  Our upper bounds show
that, under appropriate assumptions, for \emph{every} $(k,t)$-robust
equilibrium $\sigma$ in a game $\Gamma_1$ with a mediator, there exists an
equilibrium $\sigma'$ in the cheap-talk game $\Gamma_2$ corresponding to
$\Gamma_1$ that implements $\sigma$; the lower bounds in this paper show
that, if these conditions are not met, there exists a game with a
mediator and an equilibrium in that game that cannot be implemented in
the cheap-talk game.  Since our definition of games with a mediator also
allow arbitrary communication among the agents, it can also be shown that
every equilibrium in a cheap-talk game can be implemented in the
mediator game: the players simply ignore the mediator and communicate
with each other.

\subsection{Solution concepts}

We can now define the solution concepts relevant for this paper.  In
particular, we consider a number of variants of $(k,t)$ robustness,
and the motivation behind them.

In defining these solution concepts, we need to consider the expected
utility of s trategy profile conditional on players having certain types.  
We abuse notation and continue
to use $u_i$ for this, writing for example, $u_i(t_K, \sigma)$ to denote
the expected utility to player $i$ if the strategy profile $\sigma$ is
used, conditional on the players in $K$ having the types $t_K$.  Since the
strategy
$\sigma$ here can come from the underlying game or some extension of it,
the function $u_i$ is rather badly overloaded.  We sometimes include the
relevant game as an argument to $u_i$ to emphasize which game the
strategy profile $\sigma$ is taken from, writing, for example,
$u_i(t_K,\Gamma',\sigma)$.

\paragraph{$k$-resilient equilibrium:}

A strategy profile is a Nash equilibrium if no player can gain any
advantage by using a different strategy, given that all the other
players do not change their strategies. We want to define a notion
of \emph{$k$-resilient equilibrium} that generalizes Nash
equilibrium, but allows a coalition of up to $k$ players to change
their strategies.  One way of capturing this, which goes back to
Aumann~\cite{Aumann59}, is to require that there be no deviations
that result in everyone in a group of size at most $k$ doing better.

To make this intuition precise, we need some notation. Given a type
space $\T$, a set $K$ of players, and $t \in \T$, let $\T(t_K) =
\{t\,': t\,'_K = t_K\}$. If $\Gamma$ is a game over type space $\T$,
$\sigma$ is a strategy profile in $\Gamma$, and $\Pr$ is the
probability on the type space $\T$, let
$$u_i(t_K,\sigma)
= \sum_{t\,' \in \T(t_K)} \Pr(t\,' \mid \T(t_K))
u_i(t',\sigma).$$ Thus, $u_i(t_K,\sigma)$ is $i$'s expected payoff
if everyone uses strategy $\sigma$ and types are restricted to
$\T(t_K)$.

\begin{definition}\label{def:1} $\sigma$ is a \emph{$k$-resilient$'$
equilibrium} if, for all $K \subseteq N$
and all types $t \in \T$, it is not the case that there exists a
strategy $\tau$ such that
$u_i(t_K,\tau_K,\sigma_{-K}) > u_i(t_K,\sigma)$ for all $i
\in K$.
\end{definition}

Thus, $\sigma$ is
$k$-resilient$'$ if no subset $K$ of at most $k$ players can all do
better by deviating, even if they share their type information (so
that if the true type is $t$, the players in $K$ know $t_K$). 
This is essentially Aumman's~\cite{Aumann59} notion of resilience
to coalitions, except that we place a bound on the size of coalitions.

As the
prime suggests, this will not be exactly the definition we focus on.
ADGH consider a stronger notion, which requires that there be no
deviation where even one player does better.

\begin{definition}\label{def:2} $\sigma$ is a \emph{strongly $k$-resilient$'$
equilibrium} if, for all $K \subseteq N$
with $|K| \le k$
and all types $t \in \T$, it is not the case that there exists a
strategy $\tau$ such that
$u_i(t_K,\tau_K,\sigma_{-K}) > u_i(t_K,\sigma)$ for some $i \in K$.
\end{definition}

Both of these definitions have a weakness: they implicitly assume
that the coalition members cannot communicate with each other beyond
agreeing
on what strategy to use.
While, in general, there are equilibria in the cheap-talk game that
are not available in the underlying game (so having more
communication can increase the number of possible equilibria),
perhaps surprisingly, allowing
communication between coalition members can also \emph{prevent}
certain equilibria,
as the following example shows.

\begin{example} {\rm Consider a game with four players. Players 1
and 2 have a type in $\{0,1\}$; the type of players 3 and 4 is 0.
All tuples of types are equally likely.   Players 3 and 4 can each
choose an action in the set $\{0,1,\punish,\pass\}$;
players 1 and 2 each choose an action in
$\{\punish,\pass\}$.  If anyone plays
$\punish,$ then everyone gets a payoff of $-1$.  If no one plays
$\punish,$ the payoffs are as follows: If player 3 plays $\pass$,
then 3 gets a payoff of 1; similarly, if player 4 plays $\pass$, then 4
gets a payoff of 1.  If player 3
plays 0 or 1 and this is 1's type, then 3 gets 5; if not, then 3
gets -5; similarly for player 4.  Finally, if player 2's type is
$0$, then player 1 and 2's payoffs are the same
as player $3$'s payoffs; similarly, if player 2 has a 1, then 1 and
2's payoffs are the same as 4's payoffs.
It is easy to see that everyone playing $\pass$ is a 3-resilient$'$
equilibrium in the underlying game and
it is also an equilibrium in the game with a mediator, if the
players cannot communicate.   However, if players can communicate
for one round, then players 1, 2, and 3 can do better if player 1
sends player 3 his type, and player 3 plays it. This guarantees
player 3 a payoff of 5, while players 1 and 2 get an expected payoff
of $2.5$.

Now suppose that we consider a variant of this game, where the actions
are the same and the payoffs for players 3 and 4 are the same, but
the payoffs for players 1 and 2 are modified as follows.
If player 2's type is $0$, then if no one plays
$\punish$, player 1 and 2's payoffs are the same
as player $3$'s payoffs if player 4 plays $\pass$; if player 4
plays 0 or 1, then their payoff is $-5$.  Similarly, if player 2's
type is 1, then player 1 and 2's payoffs are the same player $4$'s
payoffs if player 3 plays $\pass$; if player 3 plays 0 or 1, then
their payoff is $-5$.
It is easy to show that everyone playing $\pass$ is still
3-resilient$'$ if we allow only one round of communication.  But
with two rounds of
communication, everyone playing $\pass$ is no longer
3-resilient$'$: players 1, 2, and 3 can do better if player 2 sends
player 1 his type, player 1 sends player 3 his type if player 2's
type is 0 (and sends nothing otherwise), and player 3 plays player
1's type if player 1 sends it.} \wbox
\end{example}

Since it seems reasonable to assume that coalition members will
communicate, it seems unreasonable to call everyone playing $\pass$
3-resilient if some communication among coalition members can
destroy that equilibrium. More generally, we clearly cannot hope to
implement a $k$-resilient equilibrium in the mediator game using
cheap talk if the equilibrium does not survive once we allow
communication among the coalition members. This motivates
the following definition.

\begin{definition}\label{def:3} $\sigma$ is a \emph{(strongly) $k$-resilient
equilibrium} in
a game $\Gamma'$ if, for all $K \subseteq N$ with $|K| \le k$
and all types $t \in \T$, it is not the case that there exists a
strategy $\tau$ such that
$u_i(t_K,\Gamma'+CT(K),\tau_K,\sigma_{-K}) > u_i(t_K,\Gamma',\sigma)$
for some $i
\in K$.
\end{definition}

\noindent This definition makes precise the intuition that players in
the coalition are allowed arbitrary communication among themselves.

Note that Nash equilibrium is equivalent to both 1-resilience and
strong 1-resilience; however, the notions differ for $k > 1$.
It seems reasonable in  many applications to bound the size of
coalitions; it is hard to coordinate a large coalition!  Of course,
the appropriate bound on the size of the coalition may well depend
on the utilities involved. Our interest in strong $k$-resilience was
motivated by wanting to allow situations where one player
effectively controls the coalition. This can happen in practice in a
network if one player can ``hijack'' a number of nodes in the
network.  It could also happen if one player can threaten others, or
does so much better as a result of the deviation that he persuades
other players to go along, perhaps by the promise of side payments.
While it can be argued that, if there are side payments or threats,
they should be modeled in the utilities of the game, it is sometimes
more convenient to work directly with strong resilience.
In this paper we consider
both resilience and strong resilience, since the results on
implementation obtained using the different notions are
incomparable. Just because a strongly resilient strategy in a game
with a mediator can be implemented by a strongly resilient strategy
in a cheap-talk game, it does not follow that a resilient strategy
with a mediator can be implemented by a resilient strategy in a
cheap-talk game, or vice versa.
However, as we show, we get the same lower bounds for both
resilience and strong resilience: in our lower bounds, we give games
with mediators with a strongly $k$-resilient equilibrium $\sigma$
and show that there does not exist a cheap-talk game and a strategy
that $\sigma'$ that implements $\sigma$ and is $k$-resilient.
Similarly, we can show that we get the same upper bounds with both
$k$-resilience and strong $k$-resilience.
Other notions of resilience to coalitions have been defined in the
literature.  For example,
Bernheim, Peleg, and Whinston \cite{BernheimPelegWhinston} define a
notion of \emph{coalition-proof Nash equilibrium} that, roughly
speaking, attempts to capture the intuition that $\sigma$ is a
coalition-proof equilibrium if there is no deviation that allows all
players to do better. However, they argue that this is too strong a
requirement, in that some deviations are not \emph{viable}: they are
not immune from further deviations. Thus, they give a rather
complicated definition that tries to capture the intuition of a
deviation that is immune from further deviations.  This work is
extended by Moreno and Wooders \cite{MW96} to allow correlated
strategies. Although it is beyond the scope of this paper to go
through the definitions, it is easy to see that our impossibility
results apply to them, because of the particular structure of the
games we consider.

For some of our results we will be interested in strategies that
give ``almost'' $k$-resilience, in the sense that no player in a
coalition can do more than $\epsilon$ better by
deviating, for some small $\epsilon$.

\begin{definition} If $\epsilon \ge 0$, then
$\sigma$ is an
\emph{\eps $k$-resilient
equilibrium} in a game $\Gamma'$ if, for all $K \subseteq N$
with $|K| \le k$
and all types $t \in \T$, it is not the case that there exists a
strategy $\tau$ such that
$u_i(t_K,\Gamma'+CT(K),\tau_K,\sigma_{-K}) > u_i(t_K,\Gamma',\sigma) +
\epsilon$ for all
$i \in K$.
\end{definition}
Clearly if $\epsilon = 0$, then an \eps $k$-resilient equilibrium is
a $k$-resilient equilibrium.  

\paragraph{$(k,t)$-robust equilibrium:} 
We now define the main solution concept used in this paper:
$(k,t)$-robust equilibrium.  The $k$ indicates the size of
coalition we are willing to tolerate, and the $t$ indicates the number
of players with unknown utilities.  These $t$ players are analogues of
faulty players or adversaries in the distributed computing literature,
but we can think of them as being perfectly rational.  Since we do not
know what actions these $t$ players will perform, nor do we know their
identities, we are interested in
strategies for which the payoffs of the remaining players are
immune to what the $t$ players do.

\begin{definition}\label{def:robust}
A strategy profile $\sigma$ in a game
$\Gamma$ is \emph{$t$-immune} if,
for all $T
\subseteq N$ with $|T| \leq t$, all strategy profiles $\tau$, all $i
\notin T$, and all types $t_i \in \T_i$ that occur with positive
probability, we have
$u_i(t_i,\Gamma+CT(T),\sigma_{-T},\tau_T) \ge u_i(t_i,\Gamma, \sigma)$.
\end{definition}
Intuitively, $\sigma$ is $t$-immune if there is nothing that
players in a set $T$ of size at most $t$ can do to give
the remaining players a worse payoff, even if the players in $T$
can communicate.

Our notion of $(k,t)$-robustness requires both $t$-immunity and
$k$-resilience.  In fact, it requires $k$-resilience no matter what up
to $t$ players do.  That is, we require that
no matter what $t$ players do, no subset of size at most $k$ can all
do better by deviating, even with the help of the $t$ players, and even
if all $k+t$ players share their type information.

\begin{definition}
Given $\epsilon \ge 0$,
$\sigma$ is an \emph{$\epsilon$--$(k,t)$-robust equilibrium} in game
$\Gamma$ if $\sigma$ is
$t$-immune and, for all $K, T \subseteq N$ such that
$|K| \leq k$, $|T| \leq t$, and $K \cap T = \emptyset$,
and all types $t_{K \cup T} \in \T_{K \cup T}$ that occur with positive
probability, it is not the case that
there exists a
strategy profile $\tau$ such that
$$u_i(t_{K\cup T},\Gamma+CT(K \cup T),\tau_{K \cup T},\sigma_{-(K
\cup T)}) >
u_i(t_i,\Gamma+CT(T),\tau_{T}, \sigma_{-T}) +
\epsilon$$ for all $i \in K$.
A $(k,t)$-robust equilibrium is just a 0--$(k,t)$-robust equilibrium.
\end{definition}

We can define a \emph{strongly $(k,t)$-robust equilibrium} by analogy to
the definition strongly $k$-resilient equilibrium: we simply change
``for all $i \in K$'' in the definition of $(k,t)$-robust equilibrium to
``for some $i \in K$''.  Thus, in a strongly $(k,t)$-robust equilibrium,
not even a single agent in $K$ can do better if all the players in $K$
deviate, even with the help of the players in $T$.

Note that a $(1,0)$-robust equilibrium is just a Nash equilibrium,
and an $\epsilon$--$(1,0)$-robust equilibrium is what has been called an
$\epsilon$-Nash equilibrium in the literature.  
A (strongly) $(k,0)$-robust equilibrium is just a (strongly)
$k$-resilient equilibrium.
The notion $(0,t)$-robustness is somewhat in the spirit of
Eliaz's~\cite{Eliaz00} notion of $t$ fault-tolerant implementation.
Both our notion of $(0,t)$-robustness and Eliaz's notion of
$t$-fault tolerance require that what the players not in $T$ do is a
best response to whatever the players in $T$ do (given that all the
players not in $T$ follow the recommended strategy);
however, Eliaz does not require an analogue of $t$-immunity.

In this paper, we are interested in the question of when a
$(k,t)$-robust equilibrium $\sigma$ in a game $\Gamma_d$ with a
mediator extending an underlying game $\Gamma$ can be implemented by
an $\epsilon$--$(k,t)$-robust equilibrium $\sigma'$ in the cheap-talk
extension $\Gammact$ of $\Gamma$. If this is the case, we say that
$\sigma'$ is an \emph{$\epsilon$--$(k,t)$-robust} implementation of
$\sigma$. (We sometimes say that $(\Gammact,\sigma')$ is an
\emph{$\epsilon$--$(k,t)$-robust} implementation of
$(\Gamma_d,\sigma)$ if we wish to emphasize the games.)

\section{The Possibility Results}\label{sec:upperbound}

All of our possibility results have the flavor ``if there is a
$(k,t)$-robust equilibrium in a game with a mediator, then (under the
appropriate assumptions) we can implement this equilibrium using cheap
talk.''
To state the results carefully, we must define the notions of a punishment
strategy and a utility variant.
\begin{definition}
If $\Gamma_d$ is an extension of an underlying game $\Gamma$ with a
mediator $d$, a strategy profile $\rho$ in $\Gamma$ is a
\emph{$k$-punishment strategy with respect to a strategy profile
$\sigma$ in $\Gamma_d$} if for all subsets
$K \subseteq N$ with
$|K| \leq k$, all strategies $\phi$ in
$\Gamma +CT(K)$, all types $t_K \in T_K$, and all
players $i \in K$:
$$
u_i(t_K, \Gamma_d, \sigma) > u_i(t_K, \Gamma
+CT(K), \phi_{K},\rho_{-K}).
$$
If the
inequality holds with $\ge$ replacing $>$,
$\rho$ is a \emph{weak} $k$-punishment strategy with respect to
$\sigma.$
\end{definition}

\noindent Intuitively, $\rho$ is $k$-punishment strategy with
respect to $\sigma$ if, for any coalition $K$ of at most $k$
players, even if the players in $K$ share their type information, as
long as all players not in $K$ use the punishment strategy in the
underlying game, there is nothing that the players in $K$ can do in
the underlying game that will give them a better expected payoff
than playing $\sigma$ in $\Gamma_d$.

\fullv{Notice that if $k + t < n \le 2k + t$, $\Gamma_d$ is a
mediator game extending $\Gamma$, and $\sigma$ is a $(k,t)$-robust
equilibrium in $\Gamma_d$, then there cannot be a $(k+t)$-punishment
strategy with respect to $\sigma$. For if $\sigma$ is a
$(k+t)$-punishment strategy, consider the strategy in the mediator
game where a set $T$ with $|T|= t \ge n - (k+t)$ players do not send
a message to the mediator, and just play $\rho$. If $\sigma$ is
$t$-immune, $u_i(\sigma_{N-T},\rho_T) \ge u_i(\sigma)$.  But in the
underlying game, if the players in $N-T$ share their types, they can
compute what the mediator would have said, and thus can play
$\sigma_{N-T}$, contradicting the assumption that $\sigma$ is a
punishment strategy. }

The notion of utility variant is used to make precise that certain
results do not depend on knowing the players' utilities; they hold
independently of players' utilities in
the game. A game $\Gamma\,'$ is
a \emph{utility variant} of a game $\Gamma$ if $\Gamma\,'$ and
$\Gamma$ have the same game tree, but the utilities of the players
may be different in $\Gamma$ and $\Gamma\,'$. Note that if
$\Gamma\,'$ is a
utility variant of $\Gamma$, then $\Gamma$ and $\Gamma\,'$ have the
same set of strategies.  We use the notation $\Gamma(u)$ if we want
to emphasize that $u$ is the utility function in game $\Gamma$.  We
then take $\Gamma(u\hspace{0.5mm}')$ to be the utility variant of
$\Gamma$ with utility functions $u\hspace{0.5mm}'$.

We say that \emph{broadcast can be simulated}, if for all
$\delta>0$, broadcast channels can be implemented with probability
$1-\delta$. Broadcast can be simulated if, for example, there are
broadcast channels; or if there is a trusted preprocessing phase
where players may broadcast and assuming cryptography; or if
unconditional pseudo-signatures are established \cite{PW96}.

In the theorem, we take ``assuming cryptography'' to be a shorthand
for the assumption that \emph{oblivious transfer} \cite{R81,EGL85}
can be implemented with probability $1-\epsilon$ for any desired
$\epsilon>0$.
It is known that this assumption holds if \emph{enhanced trapdoor
permutations} exist, players are computationally bounded, and the
mediator can be described by a polynomial-size circuit
\cite{goldreich03}.
\begin{theorem}\label{thm:upperbound}
Suppose that $\Gamma$ is Bayesian game with $n$ players and
utilities $u$, $d$ is a mediator that can be described by a circuit
of depth $c$, and $\sigma$ is a
$(k,t)$-robust equilibrium of a game $\Gamma_d$ with a mediator $d$.
\beginsmall{enumerate}
\item[(a)]
If $3(k+t)< n$, then there exists a strategy
$\sigmact$ in
$\Gammact(u)$ such that for all utility variants $\Gamma(u')$, if
$\sigma$ is a $(k,t)$-robust equilibrium of $\Gamma_d(u')$, then
$(\Gammact(u'),\sigmact )$ implements $(\Gamma_d(u'), \sigma)$.
The running time of $\sigmact $ is $O(c)$.

\item[(b)]
If $2k+3t < n$ and there exists a $(k+t)$-punishment strategy with
respect to $\sigma$, then there exists a strategy $\sigmact $ in
$\Gammact$ such that
$\sigmact $ implements  $\sigma$.
The expected running time of $\sigmact $ is $O(c)$.

\item[(c)]
If $2(k + t) < n$ and broadcast channels can be simulated, then, for
all $\epsilon > 0$, there exists a strategy $\sigmact ^\epsilon$
in $\Gammact$ such that
$\sigmact ^\epsilon$ $\epsilon$-implements $\sigma$.
The running time of
$\sigmact ^{\epsilon}$ is $O(c)$.

\item[(d)]
If $k + t < n$ then, assuming cryptography and
that a PKI is in place,
there
exists a strategy $\sigmact ^\epsilon$ in $\Gammact$ such that
$\sigmact ^\epsilon$ $\epsilon$-implements
$\sigma$.
The
expected
running time of $\sigmact ^{\epsilon}$
is $O(c)\cdot f(u) \cdot O(1/\epsilon)$ where $f(u)$ is a function
of the utilities.

\item[(e)]
If $k + 3t < n$ or if $k + t < n$ and
a trusted PKI is in place, and there exists a
$(k+t)$-punishment
strategy with respect to $\sigma$,
then, assuming cryptography,
there
exists a strategy $\sigmact ^\epsilon$ in $\Gammact$ such that
$\sigmact ^\epsilon$ $\epsilon$-implementers $\sigma$.
The
expected
running time of $\sigmact ^{\epsilon}$
is $O(c) \cdot f(u)$ where $f(u)$ is a function of the utilities but
is independent of $\epsilon$.

\endsmall{enumerate}
\end{theorem}

Note that in part (a) we say ``the running time'', while in the other
parts we say ``the expected running time''.  Although all the strategies
used are behavior strategies (and thus use randomization), in part (a),
the running time is bounded, independent of the randomization.  In the
remaining parts, we cannot put an \emph{a priori} bound on the running
time; it depends on the random choices.  As our lower bounds show, this
must be the case.

We briefly comment on the differences between
\theoremref{thm:upperbound} and the corresponding Theorem 4 of ADGH.
In ADGH, we were interested in finding strategies that were not only
$(k,t)$-robust, but also survived iterated deletion of weakly
dominated strategies. Here, to simplify the exposition, we just
focus on $(k,t)$-robust equilibria.   
For part (a), in ADGH, a behavioral strategy was used that had no
upper bound on running time.  This was done in order to obtain a
strategy that survived iterated deletion.  However, it is observed
in ADGH that, without this concern, a strategy with a known upper
bound can be used.  As we observed in the introduction,
part (a), as stated,
actually follows from known
results in multiparty computation
\cite{BGW88,CCD88}.  Part (b) here is the same as in ADGH.
In part (c), we assume here the ability to
simulate broadcast; ADGH assumes cryptography.  As we have observed,
in the presence of cryptography, we can simulate broadcast, so the
assumption here is weaker.  In any case, as observed in the
introduction, part (c) follows from known results \cite{RB89}.
Parts (d) and (e) are new, and will be proved in \cite{ADGH06full}.
The proof uses ideas from \cite{GMW87} on multiparty computation.  For
part (d), where there is no punishment strategy, ideas from \cite{EGL85}
on getting \emph{$\epsilon$-fair} protocols are also required.
(An $\epsilon$-fair protocols is one where
if one player knows the mediator's value with probability $p$, then other
players know it with probability at least $p-\epsilon$.)
Our proof of part (e) shows that if $n > k+3t$, then we can essentially set
up a PKI on the fly.
These results
strengthen Theorem 4(d) in ADGH, where punishment was required and $n$
was required to be greater than $k+2t$.

\section{The Impossibility Results}\label{sec:lowerbounds}

\subsection*{No bounded implementations}
We prove that it is impossible to get an implementation with bounded
running time in general if $2k + 3t < n \le 3k+3t$. This is true even if
there is a punishment strategy. This result is optimal.
If $3k + 3t < n$, then there does exist a bounded implementation;
if $2k + 3t < n \le 3k+3t$ there exists an unbounded implementation that
has constant \emph{expected} running time.
\begin{theorem}\label{thm:lowerbound-4}
If $2k + 3t < n \le 3k+3t$, there is a game $\Gamma$ and a
strong $(k,t)$-robust equilibrium $\sigma$ of a game $\Gamma_d$ with
a mediator $d$ that extends $\Gamma$ such that there exists a
$(k+t)$-punishment strategy with respect to $\sigma$ for which there
do not exist a natural number $c$ and a strategy  $\vecsigmact$ in the
cheap talk game extending $\Gamma$
such that the running time of $\vecsigmact$ on the equilibrium path is
at most $c$ and
$\vecsigmact$ is a $(k,t)$-robust implementation of $\sigma$.
\end{theorem}
\begin{proof}
We first assume that $n=3$, $k=1$, and $t=0$.  We consider a family
of 3-player games $\Gamma_3^{n,k+t}$, where $2k+3t < n \le
3k+3t$, defined as follows. Partition $\{1, \ldots, n\}$ into three sets
$B_1$, $B_2$, and $B_3$, such that
$B_1$ consists of the first $\lfloor n/3 \rfloor$ elements in
$\{1, \ldots, n\}$, $B_3$ consists of the last $\lceil n/3 \rceil$
elements, and $B_2$ consists of the remaining elements.

Let $p$ be a prime such that $p > n$.
Nature chooses a polynomial $f$
of degree $k+t$ over the $p$-element field GF($p$) uniformly at random.
For $i \in \{1,2,3\}$, player $i$'s  type
consists of the set of pairs $\{ (h, f(h)) \mid h \in B_i
\}$. Each player wants to learn $f(0)$ (the secret), but would
prefer that other players do not learn the secret. Formally, each
player must play either 0 or 1.   The utilities are
defined as follows:
\beginsmall{itemize}
\item if all players output $f(0)$ then all players get 1;
\item if player $i$ does not output $f(0)$ then he gets $-3$;
\item otherwise players $i$ gets $2$.
\endsmall{itemize}

Consider the
mediator game where each player is supposed to
tell the mediator his type.
The mediator records all the pairs $(h,v_h)$ it receives. If at
least $n-t$ pairs are received  and there exists a unique degree
$k+t$ polynomial that agrees with at least $n-t$ of the pairs then
the mediator interpolates this unique polynomial $f'$ and sends
$f'(0)$ to each player; otherwise,
the mediator sends 0 to each player.

Let $\sigma_i$ be the strategy where player $i$ truthfully tells the
mediator his type and follows the mediator's recommendation.
It is easy to see that $\sigma$ is a $(1,0)$-robust equilibrium
(i.e., a Nash equilibrium).
If a player $i$ deviates  by misrepresenting or not telling the
mediator up to $t$ of his shares, then everyone still learns; if the
player misrepresents or does not tell the mediator about more of his
shares, then the mediator sends the default value 0.
In this case $i$ is worse off. For if 0 is  indeed the secret, which
it is with probability 1/2,
$i$ gets 1 if he plays 0, and $-3$
if
he plays 1.  On the other
hand, if 1 is the secret, then $i$ gets 2 if he plays 1
and $-3$ otherwise.
Thus, no matter what $i$ does, his expected utility is at most
$-1/2$.
This argument also shows that if $\rho_i$ is
the strategy where $i$ decides 0 no matter what,  then
$\rho$ is a 1-punishment strategy with respect to~$\sigma$.

Suppose, by way of contradiction, that there is a cheap-talk strategy
$\sigma'$ in the game $\Gammact$ that implements $\sigma$ such
that any execution of $\sigma'$ takes at most $c$
rounds.
We say that a player $i$ \emph{learns the secret by round $b$ of
$\sigma'$} if, for all executions
(i.e., plays)
$r$ and $r'$ of $\sigma'$
such that $i$ has the same type and the same
message history up to round $b$,
the secret is the same in $r$ and $r'$.  Since we have assumed that all
plays of $\sigma'$ terminate in at most $c$ rounds, it must be the case
that all players learn the secret by round $c$ of $\sigma'$.  For if
not, there are two executions $r$ and $r'$ of $\sigma'$ that $i$ cannot
distinguish by round $c$,
where the secret is different in $r$ and $r'$.
Since $i$ must play the same move in $r$ and $r'$, in one case he is not
playing the secret, contradicting the assumption that $\sigma'$
implements $\sigma$.
Thus, there must exist a round $b \le c$ such that all three players
learn the secret at round $b$ of $\sigma'$ and, with nonzero
probability, some player, which we can assume without loss of generality
is player 1, does not learn the secret at round $b-1$ of $\sigma'$.
This means that there exists
a type $t_1$ and message history $h_1$ for player $1$
of length $b-1$
that occurs with
positive probability when player 1 has type $t_1$ such that, after $b-1$
rounds, if player 1 has type $t_1$ and history $h_1$, player $1$
considers it possible that the secret could be either
0 or 1.
Thus, there must exist type profiles $t$ and $t'$ that correspond to
polynomials $f$ and $f'$ such that $t_1=t'_1$, $f(0)\neq f'(0)$ and,
with positive probability, player 1 can have history $h_1$ with both
$t$ and $t'$, given that all three players play $\sigma'$.

Let $h_2$ be a history for player 2
of length $b-1$
compatible with $t$ and $h_1$
(i.e.,
when the players play $\sigma'$,
with positive probability,
player 1 has $h_1$, player 2 has
$h_2$, and the true type profile is $t$); similarly, let $h_3$ be a
history
of length $b-1$
for player 3 compatible with $t'$ and $h_1$.
Note that player $i$'s action according to $\sigma_i$ is
completely determined by his type, his message history, and the outcome
of his coin tosses.
Let $\sigma'_2[t_2,h_2]$
be the strategy for player 2
according to which player 2 uses $\sigma'_2$ for
the first $b-1$ rounds, and then from round $b$ on,
player 2 does what it would have done according to $\sigma_2'$
if its type had been $t_2$ and its message history for the
first $b-1$ rounds had been $h_2$ (that is, player 2 modifies his actual
message history by replacing the
prefix
of length $b-1$ by $h_2$, and leaving
the rest of the message history unchanged).  We
can similarly define $\sigma'_3[t_3',h_3]$. Consider the strategy
profile $(\sigma'_1, \sigma'_2[t_2,h_2], \sigma'_3[t_3',h_3])$.
Since $\sigma'_i[t_i,h_i]$ is identical to $\sigma'_i$ for the first
$b-1$ steps, for $i = 2,3$, there is a positive probability that
player 1 will have history $h_1$ and type $t_1$ when this strategy
profile is played.  It should be clear that, conditional on this
happening, the probability that player 1 plays 0 or 1 is independent
of the actual types and histories of players 2 and 3. This is
because players 2 and 3's messages from time $b$ depend only on
$i$'s messages, and not on their actual type and history.  Thus, for
at least one of $0$ and $1$, it must be the case that the
probability that player 1 plays this value is strictly less than 1.
Suppose without loss of generality that the probability of
playing
$f(0)$ is less than 1.

We now claim that $\sigma'_3[t_3',h_3]$ is a profitable
deviation for player 3.  Notice that player 3 receives the same messages
for the first $b$ rounds of $\sigma'$ and $(\sigma'_1, \sigma'_2,
\sigma'_3[t_3',h_3])$.  Thus, player 3 correctly plays the secret no
matter what the type profile is, and gets payoff of at least 1.
Moreover, if the type profile is $t$,
then, by construction, with positive probability, after $b-1$ steps,
player 1's history will be $h_1$ and player 2's history will be $h_2$.
In this case, $\sigma'_2$ is identical to $\sigma'_2[t_2,h_2]$, so the
play will be identical to $(\sigma'_1, \sigma'_2[t_2,h_2],
\sigma'_3[t_3',h_3])$.  Thus, with positive probability,
player 1
will not output $f(0)$, and player 3 will get payoff 2.  This means
player 3's expected utility is greater than 1.

For the general case, suppose that $2k+3t < n \le 3k+3t$.  Consider
the $n$-player
game $\Gamma^{n,k,t}$, defined as follows. Partition the players
into three groups, $B_0$, $B_1$, and $B_2$, as above. As in the
3-player game, nature
chooses a polynomial $f$ of degree $k+t$ over the field $\{0,1\}$
uniformly at random, but now player $i$'s type is just the pair
$(i,f(i))$. Again, the players want to learn $f(0)$, but would
prefer that other players do not learn the secret, and must output a
value in $F$.
The payoffs are similar in spirit to the 3-player game:
\beginsmall{itemize}
\item if at least $n-t$ players output $f(0)$ then all players
that output $f(0)$
get 1;
\item if player $i$ does not output $f(0)$ then he gets $-3$;
\item otherwise player $i$ gets 2.
\endsmall{itemize}

The mediator's strategy is essentially identical to that in the 3-player
game (even though now it is getting one pair $(h,v_h)$ from each player
rather than a set of such pairs from a single player).  Similarly, each
player $i$'s strategy in $\Gamma^{n,k,t}_d$, which we denote $\sigma^n_i$,
is essentially identical to the strategy in the 3-player game with the
mediator. Again, if $\rho_i^n$ is the strategy in the $n$-player game
where $i$ plays 0 no matter what his type, then it is easy to check
that $\rho^n$ is a $(k+t)$-punishment strategy with respect to
$\sigma^n$.

Now suppose, by way of contradiction, that there exists a strategy
$\sigma'$ in the cheap-talk extension $\Gammact^{n,k,t}$ of
$\Gamma^{n,k,t}$ that is a $(k,t)$-robust implementation of
$\sigma^n$
such that all executions of $\sigma'$ take at most $c$ rounds.
We show
in \appendixref{sec:proof-thm-4}
that we can use $\sigma'$
to get a
$(1,0)$-robust implementation in the 3-player mediator game
$\Gamma_{3,d}^{n,k+t}$, contradicting the argument above.
\end{proof}

\subsection*{Byzantine Agreement and Game Theory}

In \cite{ADGH06} it is shown that if $n > 3k+3t$, we can implement a
mediator in a way that does not depend on utilities and does not need a
punishment strategy.  Using novel connections to randomized Byzantine
agreement lower bounds, we show that neither of these properties hold in
general if $n \le 3k + 3t$.

We start by showing that we cannot handle all utilities variants if
$n \le 3k+3t$. Our proof exposes a new connection between utility variants and the problem of \emph{Weak Byzantine Agreement}~\cite{L83}.
Lamport~\cite{L83} showed that there is no
deterministic protocol with bounded running time for \emph{weak Byzantine
agreement} if $t \ge n/3$.  We prove a stronger lower bound for any randomized protocol that only assumes that the running time has finite expectation.

\begin{proposition}\label{lem:dist-lower-bound}
If $\max\{2,k+t\} < n\le3k+3t$, all $2^n$ input values are equally
likely, and $P$ is a (possibly randomized) protocol with finite
expected running
time (that is, for all protocols $P''$ and sets $|T| \le k+t$, the
expected  running time of processes $P_{N-T}$ given
$(P_{N-T},P''_T)$ is finite), then there exists a protocol $P'$ and
a set $T$ of players with $|T| \le k+t$ such that an execution of
$(P_{N-T},P'_T)$ is unsuccessful for the weak Byzantine agreement
problem with nonzero probability.%
\end{proposition}

\fullv{
\begin{proof}
See \appendixref{sec:proof-thm-2}.
\end{proof}
}

The idea of our impossibility result is to construct a game that
captures weak Byzantine agreement.
The challenge in the proof is that, while in the Byzantine agreement
problem, nature
chooses which processes are faulty, in the game, the players decide
whether or not to behave in a faulty way.
Thus, we must set up the incentives so
that players gain by choosing to be faulty iff Byzantine agreement
cannot be attained,
while ensuring that a $(k,t)$-robust cheap-talk implementation of the
mediator's strategy in the game will solve Byzantine agreement.

\def\lowerboundII {If $2k + 2t < n \le 3k+3t$, there is a game
$\Gamma(u)$ and a
strong $(k,t)$-robust equilibrium $\sigma$ of a game
$\Gamma_d$ with a mediator $d$ that extends $\Gamma$ such that there
exists a $(k+t)$-punishment strategy with respect to $\sigma$ and
there does not exist a strategy $\vecsigmact$ such that for all
utility variants $\Gamma(u\hspace{0.5mm}')$ of $\Gamma(u)$, if
$\sigma$ is a $(k,t)$-robust equilibrium of
$\Gamma_d(u\hspace{0.5mm}')$, then
$(\Gammact(u\hspace{0.5mm}'),\vecsigmact)$ is a $(k,t)$-robust
implementation of $(\Gamma_d(u\hspace{0.5mm}'), \sigma)$. }

\begin{theorem}\label{thm:lowerbound-2}
\lowerboundII
\end{theorem}

\fullv{
\begin{proof}
See \appendixref{sec:proof-thm-2}.
\end{proof}
}

\noindent
\theoremref{thm:lowerbound-2}
shows that
we cannot, in general, get a
\emph{uniform} implementation if $n \le 3k+3t$.  As shown in
\theoremref{thm:upperbound}(b)--(e), we can implement mediators if $n
\le 3k+3t$ by taking advantage of knowing the players' utilities.

We next prove that if $2k + 3t < n \le 3k+3t$, although mediators
can be implemented, they cannot be implemented without a punishment
strategy. In fact we prove
that they cannot even be $\eps$implemented without a punishment strategy.
Barany~\citeyear{Barany92} proves a weaker version of a special case of this result, where
$n=3$, $k=1$, and $t=0$.%
It is not clear how to extend Barany's
argument to the general case, or to $\eps$implementation. We use the power of randomized Byzantine agreement lower bounds for this result.

\def\lowerboundIII{If $2k+2t < n \le 3k + 3t$, then there exists a game $\Gamma$, an
$\epsilon > 0$, and a  strong $(k,t)$-robust equilibrium $\sigma$ of
a game $\Gamma_d$ with a mediator $d$ that extends $\Gamma$, for
which there does not exist a strategy $\vecsigmact$ in the CT game
that extends $\Gamma$ such that $\vecsigmact$ is an \eps
$(k,t)$-robust implementation of $\sigma$. }

\begin{theorem}\label{thm:lowerbound-3}
\lowerboundIII
\end{theorem}

\fullv{
\begin{proof}
See \appendixref{sec:proof-thm-3}.
\end{proof}
}

We now show that the assumption that
$n > 2k + 3t$ in \theoremref{thm:upperbound}
is necessary. More precisely, we show that
if $n \le 2k+3t$, then
there is a game with a mediator that has a
$(k,t)$-robust equilibrium
that does not have
a
$(k,t)$-robust
implementation in a cheap-talk game. We actually prove a stronger
result: we show that there cannot even be an
\eps $(k,t)$-robust implementation, for sufficiently small
$\epsilon$.

\begin{theorem}\label{thm:lowerboundb} If $k+2t < n \le 2k+3t$,
there exists a game $\Gamma$, a
strong $(k,t)$-robust
equilibrium $\sigma$ of a game $\Gamma_d$ with a mediator $d$ that
extends $\Gamma$, a $(k+t)$-punishment strategy with respect to
$\sigma$, and an $\epsilon > 0$, such that there does not exist a
strategy $\vecsigmact$ in the CT extension of $\Gamma$ such that
$\vecsigmact$ is an \eps $(k,t)$-robust  implementation of $\sigma$.
\end{theorem}

The proof of \theoremref{thm:lowerboundb} splits into two cases: (1)
$2k+2t < n \le 2k+3t$ and $t \ge 1$ and (2)
$k + 2t < n \le 2k + 2t$. For the first case, we use a reduction to
a generalization of the Byzantine agreement problem called the
$(k,t)$-\emph{\dagree problem}. This problem is
closely related to the problem of \emph{broadcast with extended
consistency} introduced by Fitzi~et al.~\citeyear{FHHW03}.

\def\lowerboundVa{If $2k + 2t < n \le 2k + 3t$ and $t\ge 1$,
there exists a game $\Gamma$, an $\epsilon > 0$, a
strong
$(k,t)$-robust equilibrium $\sigma$ of a game $\Gamma_d$ with a
mediator $d$ that extends $\Gamma$, and a $(k+t)$-punishment
strategy with respect to $\sigma$, such that there does not exist a
strategy $\vecsigmact$ in the CT extension of $\Gamma$ which is an
\eps $(k,t)$-robust  implementation of $\sigma$. }

\begin{theorem}\label{thm:lowerbound-5a}
\lowerboundVa
\end{theorem}

\fullv{
\begin{proof}
See \appendixref{sec:proof-thm-5a}.
\end{proof}
}

We then consider the second case of \theoremref{thm:lowerboundb},
where $k + 2t < n \le 2k + 2t$. Since we do not assume players know
when other players have decided
in the underlying game, our proof is a
strengthening of the lower bounds of \cite{SRA81,Heller05}.

\begin{theorem}\label{pro:lowerboundbii}
If $k + 2t < n \leq 2k+2t$, there exist a game $\Gamma$, an
$\epsilon > 0$, a mediator game $\Gamma_d$ extending $\Gamma$, a
strong $(k,t)$-robust equilibrium $\sigma$ of $\Gamma_d$, and a
$(k+t)$-punishment strategy $\rho$ with respect to $\sigma,$ such
that there is no strategy $\vecsigmact$ that is
an
\eps
$(k,t)$-robust implementation of $\sigma$ in the cheap-talk
extension of $\Gamma$, even with broadcast channels.
\end{theorem}

\fullv{
\begin{proof}
See \appendixref{sec:-proof-pro:lowerboundbii}.
\end{proof}
}

Our last lower bound using Byzantine agrement impossibilities gives tight bounds to the result of
\theoremref{thm:upperbound}(e)
for the case that $n > k+3t$.  We show that
a PKI cannot be set up on the fly if $n \le k+3t$.
Our proof is based on a reduction to a lower bound for the
\emph{$(k,t)$-partial broadcast problem},
a novel variant of Byzantine agreement that can be viewed as capturing
minimal conditions that still allow us to prove strong randomized lower
bounds.

\def\lowerboundX{If $\max(2,k+t) < n \le k + 3t$, then there is a game $\Gamma$, a
strong $(k,t)$-robust equilibrium $\sigma$ of a game
$\Gamma_d$ with a mediator $d$ that extends $\Gamma$ for which there
does not exist a strategy $\vecsigmact$ in the CT game that extends
$\Gamma$ such that $\vecsigmact$ is an \eps $(k,t)$-robust
implementation of $\sigma$ even if players are computationally
bounded and we assume cryptography. }

\begin{theorem}\label{thm:d-broadcast-lowerbound}
\lowerboundX
\end{theorem}

\fullv{
\begin{proof}
See \appendixref{sec:proof-d-broadcast-lowerbound}.
\end{proof}
}

\subsection*{Tight bounds on running time}

We now turn our attention to running times. We provide tight bounds on
the number of rounds needed to $\eps$implement equilibrium when $k + t <
n \le 2(k + t)$. When $2(k+t)<n$ then the expected running time is
independent of the game utilities and independent of $\epsilon$. We show
that for $k + t < n \le 2(k + t)$ this is not the case. The expected
running time must depend on the utilities, and if punishment does not
exist then the running time must also depend on $\epsilon$. 
\begin{theorem}\label{thm:d-lowerbound}
If $k + t < n \le 2(k + t) $ and $k \ge 1$,
then there exists a game  $\Gamma$, a mediator game $\Gamma_d$ that
extends $\Gamma$, a strategy $\sigma$ in $\Gamma_d$, and a strategy
$\rho$ in $\Gamma$ such that
\begin{itemize}
\item[(a)] for all $\epsilon$ and $b$, there exists a
utility function
$u^{\,b,\epsilon}$  such that $\sigma$ is a
$(k,t)$-robust
equilibrium in $\Gamma_d(u^{\,b,\epsilon})$ for all $b$ and
$\epsilon$, $\rho$ is  a $(k,t)$-punishment strategy with respect to
$\sigma$ in $\Gamma(u^{\,b,\epsilon})$ if $n
> k + 2t$,  and there does not exist an \eps $(k,t)$-robust
implementation of $\sigma$ that runs in expected time $b$ in the
cheap-talk extension
$\Gammact(u^{b, \epsilon})$ of $\Gamma(u^{b, \epsilon})$;

\item[(b)] there exists a utility function
$u$ such that $\sigma$ is a
$(k,t)$-robust equilibrium in
$\Gamma_d(u)$ and, for all $b$, there exists $\epsilon$ such that
there does not exist an \eps $(k,t)$-robust implementation of
$\sigma^i$ that runs in expected time $b$ in the cheap-talk
extension $\Gammact(u)$ of $\Gamma(u)$.
\end{itemize}
This is true even if players are computationally bounded, we assume
cryptography and there are broadcast channels.
\end{theorem}
\fullv{
\begin{proof}
See \appendixref{sec:proof-d-lowerbound}.
\end{proof}
}

Note that, in part (b), it is not assumed that there is a
$(k,t)$-punishment strategy with respect to $\sigma$ in
$\Gamma(u)$.  With a punishment strategy, for a fixed family of
utility functions, we can implement an \eps $(k,t)$-robust strategy
in the mediator game using cheap talk with running time that is
independent of $\epsilon$; with no punishment strategy, the running
time depends on $\epsilon$ in general.

\fullv{
\section{Conclusions}\label{sec:discussion}

We have provided conditions under which a $(k,t)$-robust
equilibrium with a mediator can be implemented using cheap talk, and
proved essentially matching lower bounds.   There are still a few gaps in
our theorems,
as well as other related issues to explore.  We list some of them here.
\begin{itemize}
\commentout{
\item What happens with broadcast?
\item Determinism?

}
\item In \theoremref{thm:upperbound}(c), we get only an
$\epsilon$-implementation for some $\epsilon > 0$.  Can we take
$\epsilon = 0$ here?
\item We require that the cheap-talk implementation be only a Nash
equilibrium.  But when we use a punishment strategy, this may
require a player to do something that results in him being much
worse off (although in equilibrium this will never occur, since if
everyone follows the recommended strategy, there will never be a
need for punishment).  It may be more reasonable to require that the
cheap-talk implementation be a \emph{sequential equilibrium}
\cite{KW82} where, intuitively, a player is making a best response
even off the equilibrium path.  To ensure that the cheap-talk
strategy is a sequential equilibrium, Ben-Porath \cite{Bp03}
requires that the punishment strategy itself be a Nash equilibrium.
We believe for our results where a punishment strategy is not
required, the cheap-talk strategy is in fact a sequential
equilibrium and, in the cases where a punishment strategy is
required, if we assume that the punishment strategy is a Nash
equilibrium, then the cheap-talk strategy will be a sequential
equilibrium.  However, we have not checked this carefully.  It would
also be interesting to consider the extent to which the cheap-talk
strategy satisfies other refinements of Nash equilibrium, such as
\emph{perfect equilibrium} \cite{Selten75}.

\item Our focus in this paper has been the synchronous case.
We are currently exploring the asynchronous case.  While we had originally
assumed that implementation would not be possible in the
asynchronous case, it now seems that many of the ideas of our
possibility results carry over to the asynchronous case.  However, a
number of new issues arise.  In particular, we need to be careful in
dealing with uncertainty. The traditional assumption in game theory
is to quantify all uncertainty probabilistically.  But with
asynchrony, part of the uncertainty involves, how long it will take
a message to arrive and when agents will be scheduled to move.  (In
general, in an asynchronous setting, one player can make many moves
before a second agent makes a single move.)   It is far from clear
what an appropriate distribution would be to characterize this
uncertainty.  Thus, the tradition in distributed computing has been
to assume that an adversary decides message delivery time and when
agents are scheduled.  The results in the asynchronous case depend
on how we deal with the uncertainty, which in turn affects the
notion of equilibrium.
\item We have assumed that in the cheap-talk game, every player can talk
directly to every other player.  It would be interesting to examine what
happens if there is a communication network which characterizes which
players a given player can talk to directly.
\item In the definition of $t$-immunity and $(k,t)$-robustness, we have
allowed 
the players in $T$ to use arbitrary strategies.  In practice, we
may be interested only in restricting each player $i$ in $T$ to using a
strategy in some predetermined set $S_i'$.
\end{itemize}
We hope to return to all these issues in future
work.

}

\newpage

\appendix

\noindent {\Huge \bf Appendix}

\section{Proofs}
This section includes the 
proofs for all results
stated in the main text.
We
repeat the statement of the results for the readers' convenience.

\subsection{Proof of
\theoremref{thm:lowerbound-2}}\label{sec:proof-thm-2}

In the weak Byzantine agreement problem, there are $n$ processes, up
to $t$ of which may be faulty (``Byzantine''). Each process has some
initial value, either $0$ or $1$.  Some processes (chosen by nature)
are faulty; their ``intention'' is to try to prevent agreement among
the remaining processes.  Each non-Byzantine process must decide $0$
or $1$. An execution of a protocol $P$ is \emph{successful for weak
Byzantine agreement} if the following two conditions hold:
\begin{enumerate}
\item [I.] (Agreement:) All the non-Byzantine processes decide on the
same value in $\{0,1\}$.
\item [II.] (Weak Nontriviality:) If all processes are non-Byzantine and
all processes have the same initial value $i$, then all the
processes must decide $i$.
\end{enumerate}

\repro{lem:dist-lower-bound} If $\max\{2,k+t\} < n\le3k+3t$, all
$2^n$ input values are equally likely, and $P$ is a (possibly
randomized) protocol with finite expected running
time, then there exists a protocol $P'$ and a set $T$ of
players with $|T| \le k+t$ such that an execution of
$(P_{N-T},P'_T)$ is unsuccessful for the weak Byzantine agreement
problem with nonzero probability. \erepro

\begin{proof}
The proof is based on the argument of \cite{FLM86}. Partition the
processes $N=\{1,\dots,n\}$ into three sets $B_0,B_1,B_2$ such that
$|B_i| \leq k+t$. Let $r_1,\dots,r_n$ be the random tapes such that
process $i$ uses tape $r_i$.

Let $c$ be an integer parameter that will be fixed later and
consider the scenario consisting of $2cn$ processes arranged into
$6cn$ sets $A_0,A_1,\dots,A_{6cn-1}$. The number of processes in the
set $A_i$ is $|B_{i \pmod 3}|$, and the \emph{indexes} of processes
$A_i$ correspond to the indexes of processes in the set $B_{i \pmod
3}$. Thus, for each value $j$ in $N$, there are $2c$ processes whose
index is set to $j$. If $j \in B_i$ then there is exactly one such
process in each set $A_{3\ell+i}$ for $\ell \in \{0,1,\dots,
2c-1\}$.

Each process whose index is $j \in N$ executes protocol $P_j$ with
random tape $r_j$.  Messages sent by processes in $A_i$ according to
$P$ reach the appropriate recipients in $A_{i-1
\pmod{6cn}},A_i,A_{i+1 \pmod{6cn}}$; the processes in $A_i$ start
with 1 if $ -6cn/4 \pmod{6cn} <i \le 6cn/4 \pmod{6cn}$ and 0
otherwise.

Any two consecutive sets $A_i,A_{i +1 \pmod{6cn}}$ define a possible
scenario denoted $S_i$ where the non-faulty processes, $A_i,A_{i +1
\pmod{6cn}}$, execute $P$ and the faulty processes simulate the
execution of all the remaining sets of processes. Let $e_i$ denote
the probability that protocol $P$ fails the weak Byzantine agreement
conditions in scenario $S_i$.

Fix $c=2^{5}b$ and consider the scenarios $S_1$ and $S_{3cn}$. Since
the expected running time is at most $b$ then by Markov inequality,
with probability at least
$7/8$ the processes in $S_1$ require at
most $8b$ rounds. So with probability at least $1/2$ both the
processes in $S_1$ and the processes in $S_{3cn}$ decide in at most
$8b$ rounds, denote this event as $\mathcal{E}$.

Since we fixed $c=2^{5}b$ and all processes that have the same index
use the same random tape then given $\mathcal{E}$, we claim that the
processes in $S_1$ cannot distinguish their execution from an
execution where all processes are non-faulty and begin with a 1 and
similarly processes in $S_{3cn}$ cannot distinguish their execution
from an execution where all processes are non-faulty and begin with
a 0. Therefore, given $\mathcal{E}$, the non-faulty processes in
$S_1$ decide $1$ and the non-faulty processes in $S_{3cn}$ decide
$0$. Given  $\mathcal{E}$, it cannot be the case that $e_i=0$ for
all $i$.
Hence there must exist an index $i$ such that $e_i >0$
given  $\mathcal{E}$.

Now consider the strategy $P'$ for processes in $B_{i-1 \pmod 3}$
that guesses the random tapes of the processes in $B_i,B_{i+1 \pmod
3}$ and simulates the processes $A_0,A_1,\dots,A_{6cn-1}$ except for
the processes in
$A_i,A_{i+1 \pmod{6cn}}$. With non-zero probability, the
initial values of the non-faulty processes will be as the initial
values of
$A_i,A_{i+1 \pmod{6cn}}$. Given this, with probability
$1/2$ event $\mathcal{E}$ occurs. Given this, the faulty processes
may guess the tapes of the non-faulty processes with non-zero
probability. Given this, with probability $e_i>0$ the non-faulty
processes will fail.
\end{proof}

\rethm{thm:lowerbound-2}
\lowerboundII
\erethm

\begin{proof}

Consider the following game $\Gamma$ with $2k + 2t <  n\le 3k+3t$
players. A player's type is his initial value, which is either 0 or
1. We assume that each of the $2^n$ tuples of types is equally
likely. Each player must choose a characteristic G (for ``good'') or
B (for ``bad''); if the player chooses G, he must also output either
0, 1, or $\punish$.  The utility function $u$ is characterized as
follows:
\beginsmall{itemize}
\item If there exists a set $R$ of at least $n - (k+t)$ players that
choose G such that
\beginsmall{itemize}
\item all players in $R$ either commonly output 0 or commonly output 1; and
\item if all players in $R$ have initial value $i$ then the common
output of $R$ is $i$;
\endsmall{itemize}
then we call this a \emph{good outcome}, the utility is $1$ for
players that choose G and output the same value as the players in
$R$, and the utility is 0 for all other players.

\item If $n-(k+t)$ or more players choose G and output $\punish$
or all players choose B, then we call this a \emph{punishment
outcome}; the utility is $-1$ for all players.

\item Otherwise
we have a \emph{bad outcome}, and the utility is $-2n$ for players
that choose G and $2$ for players that choose B.
\endsmall{itemize}
For future reference, we take the utility function $u^M$ to be
identical to $u$, except that in a punishment outcome, a player that
chooses B gets $2M$, while a player that chooses G gets $-2nM$ (so
that $u = u^1$).

Consider the strategy $\sigma_i$ for player $i$ in the game
$\Gamma_d$ with a mediator based on $\Gamma$ where $i$ sends its
value to the mediator, and chooses characteristic G and outputs the
value the mediator sends
if the value is in $\{0,1\}$, and outputs 0 if the mediator sends
{\punish}. The mediator sends {\punish} if there are less than
$n-(k+t)$ values sent; otherwise it sends the majority value (in
case of a tie, it sends $1$).  Let $\rho_i$ be the strategy in the
underlying game $\Gamma$ of choosing B and outputting 0.

\begin{lemma}\label{lem:lowerbound-2}
The strategy $\sigma$ is a strong $(k,t)$-robust equilibrium in the
utility variant game $\Gamma_d(u^M)$ for all $M$. Moreover, $\sigma$
results in a good outcome and $\rho$ is a $(k+t)$-punishment
strategy with respect to $\sigma$.
\end{lemma}

\begin{proof}

If $|T| \le t$, and all the players in $N-T$ play $\sigma$, then the
mediator will get at least $n-t$ values.  If the majority value is
$i$, then all the players in $N-T$ will decide $i$. Since $n > 2t$,
there must be at least one player in $N-T$ that has type $i$.
Moreover, $|N-T| \ge n-(k+t)$, so the players in $N-T$ constitute a
set $R$ that makes the outcome good.  Thus we have  $t$-immunity.

Now fix $K,T \subseteq N$ such that $K,T$ are disjoint, $|K| \leq
k$, and $|T| \leq t$. Clearly for any $\tau_{K\cup T} \in S_{K\cup
T}$, and $i \notin K \cup T$ we have
$u_i(\sigma_{-(K \cup T)}, \tau_{(K \cup T)})=1$. If at least
$n-(k+t)$ players play $\sigma$ then the outcome will be good and
all the players that play $\sigma$ will get a utility of 1, no
matter what the other players do; moreover, any player that chooses
B will get utility of 0. Here we need the fact that $n > 2k + 2t$,
so there cannot be two sets of size at least $n-(k+t)$ where the
players output different values. It easily follows that $\sigma$ is
a $(k,t)$-robust equilibrium. Note that if any set of $n-(k+t)$ or
more players play $\rho$ in the underlying game $\Gamma$ then, no
matter what the remaining players do, the utility for all the
players is $-1$, whereas, as we have seen, if $n-(k+t)$ or more
players play $\sigma$ in $\Gamma_d$, then these players get 1. Thus,
$\sigma$ is a $(k+t)$-punishment strategy with respect to $\sigma$.
\end{proof}

Returning to the proof of
\theoremref{thm:lowerbound-2},
by way of
contradiction, suppose that there exists a strategy $\sigma'$ in the
CT extension $\Gammact$ of $\Gamma$ such that
$(\Gammact(u^M),\sigma')$ is a $(k,t)$-robust implementation of
$(\Gamma_d(u^M), \sigma)$, for all $M$.
Let $P_i $ be the
protocol for process $i$ where process $i$ sends messages according
to $\sigma'_i$, taking its initial value to be its type, decides
$\ell$ if $\sigma'_i$
chooses G and outputs $\ell \in \{0,1\}$,
and outputs 0 if $i$ chooses B or chooses G and outputs $\punish$.

By \propref{lem:dist-lower-bound}, there exists a protocol $P'$,
sets $K, T$ with $|K| \le k$ and $|T| \le T$, and an $\epsilon > 0$ such that
$(P_{-(K \cup T)}, P'_{K \cup T})$ has probability $\epsilon$ of having an
unsuccessful execution.
Without loss of generality, we can assume that $|K| = k$ and $|T| = t$.
(If not, we can
just add $k-|K|$ processes to $K$ and $t - |T|$ processes to $T$ and
have them all use
protocol $P$.)
Let $\sigma_j''$ be the strategy where
player $j$ chooses $B$ and sends messages according to $P'_j$. It is
easy to see that
$(\sigma'_{-(K \cup T)},\sigma''_{K \cup T})$ results in a bad outcome
whenever $(P_{-(K \cup T)},P'_{K \cup T})$ results in an unsuccessful outcome.
For if $(\sigma'_{-(K \cup T)},\sigma''_{K \cup T})$ results in a
punishment outcome,
then all players not in $K \cup T$ output 0 with $(P_{-(K \cup T)},P'_{K
\cup T})$, so the
outcome is successful.
Thus, the probability of a bad outcome with
$(\sigma'_{-(K \cup T)},\sigma''_{K \cup T})$ is $\epsilon$.

Fix $M>
2/\epsilon$.  In the game $\Gammact(u^M)$, if $j \in K$, we
have $u^M_j(\sigma'_{-(K \cup T)},\sigma''_{K \cup T}) > 3$ (since $j$'s
expected utility
conditional on a bad outcome is greater than $4/\epsilon$, and a bad
outcome occurs with probability $\epsilon$, while $j$'s expected utility
conditional on a good or punishment outcome is at least $-1$).
Since $\sigma'$ is a $(k,t)$--robust
equilibrium, if $j \in K$,
we must have $u_j^M(\sigma'_{-T},\sigma''_{T})
\ge u_j^M(\sigma'_{-(K \cup T)},\sigma''_{K \cup T}) > 3$.
Thus, the probability of a bad outcome with
$(\sigma'_{-T},\sigma''_T)$ must be positive. Note that
$t$-immunity guarantees that, for all
$i \notin T$,
$u_i^M(\sigma'_{-T},\sigma''_{T}) \ge 1$. Thus, the total
expected utility of the players in $N-T$ when playing
$(\sigma'_{-T},\sigma''_{T})$ must be at least $n-t+2k$
(since the players in $K$ have expected utility at least 3).
However,
in a good outcome, their total utility $n-(t+k)$; in a punishment
outcome, their total utility is $-n +t < 0$; and in a bad
outcome, their total utility is less than 0 (since even if all but
one of the players in $N-(T-K)$ choose characteristic B and get
utility $2M$, the player who chooses characteristic G gets utility
$-2Mn$).
Thus, the only way that the total expected utility of the players in
$N-T$ can be greater than $n-t +2k$ is if $k=0$ and the probability
of a bad outcome (or a punishment outcome) with
$(\sigma'_{-(K \cup T)},\sigma''_{K \cup T})$ is 0.
This gives us the desired
contradiction, and completes the proof of
\theoremref{thm:lowerbound-2}.
\end{proof}

\subsection{Proof of
\theoremref{thm:lowerbound-3}}\label{sec:proof-thm-3}

\rethm{thm:lowerbound-3}
\lowerboundIII
\erethm

\begin{proof}
Consider a variant of the game described in the proof of
\theoremref{thm:lowerbound-2}:

Game $\Gamma$ has $2k + 2t <  n\le 3k+3t$ players. Players are
partitioned into three sets $B_1,B_2,B_3$ such that $|B_i| \le k+t$.
Nature chooses three independent uniformly random bits $b_1,b_2,b_3
\in \{0,1\}$ and gives each player in $B_i$ the type $b_i$.
Each player must choose a characteristic G or B; if the player
chooses G, he must also output either 0 or 1.  The utility function
$u$ is characterized as follows:
\beginsmall{itemize}
\item If there exists a set $R$ of at least $n - (k+t)$ players that
choose G such that
\beginsmall{itemize}
\item all players in $R$ either commonly output 0 or commonly output 1; and
\item if all players in $R$ have initial value $i$ then the common
output of $R$ is $i$;
\endsmall{itemize}
then we call this a \emph{good outcome}, the utility is $1$ for
players that choose G and output the same value as the players in
$R$, and the utility is 0 for all other players.

\item Otherwise we have a \emph{bad outcome}, and the utility is $0$ for players
that choose G and $16$ for players that choose B.
\endsmall{itemize}

Consider the same mediator as in \theoremref{thm:lowerbound-2},
except that rather than sending {{\punish}} if fewer than $n-(k+t)$
values are sent, the mediator simply sends 0. Again, it is easy to
see that the strategy $\sigma$ of sending the true type and
following the mediator's advice is a $(k,t)$-robust equilibrium in
the mediator game.
Note that there is no $(k+t)$-punishment strategy with respect to
$\sigma$ in this game.

Let $\sigma'$ be any strategy in the cheap talk game $\Gammact$
such that for any set $K \cup T$ with $|K\cup T| \le k+t$ and any
protocol $\tau_{K \cup T}$ the expected running time of
$(\sigma'_{N-(K \cup T)}, \tau_{K \cup T})$ is finite. Let $P$ be
the protocol for Byzantine agreement induced by $\sigma'$.
Specifically, protocol $P_i$ simulates $\sigma'_i$ by giving it its
initial value,
sending messages according to $\sigma'_i$ and finally
decide on the same value that $\sigma'_i$ outputs.

We use the following lower bound on randomized Byzantine agreement
protocols.

\begin{proposition}\label{prop:Byz-lower}
If $2t < n \le 3t$ and processes are partitioned into three sets
$B_1,B_2,B_3$ such that $|B_i| \le t$. Nature chooses three
independent uniformly random bits $b_1,b_2,b_3 \in \{0,1\}$ and
gives each player in $B_i$ the initial value $b_i$.
Then there exists a function $\Psi$ that maps protocols to protocols
 such that for any joint
protocol $P$ there exists a set $T$ of processes such that $T = B_i$
for some $i \in \{1,2,3\}$ and the execution $(P_{N-T}, \Psi(P)_T)$
fails the Byzantine Agrement problem with probability at least 1/6.
The running time of $\Psi(P)$ is polynomial in the number of players
and the running time of $P$.
\end{proposition}

\begin{proof}
The proof follows from \cite{KY84}. See \propref{prop:dagree} for a
self contained proof that also handles this special case.
\end{proof}

Let $T$ be the set whose existence is guaranteed by
\propref{prop:Byz-lower} for protocol $P$. Then consider the
strategy $\tau_T$ in $\Gammact$ where players choose B and play
according to the protocol $\Psi(P)_T$. Since with probability at
least $1/6$ the execution of $(P_{N-T},\Psi(P)_T)$ fails the
Byzantine Agrement problem then with probability at least $1/6$ the
execution of $(\sigma'_{N-T}, \tau_T)$ reaches a bad outcome and the
expected utility of each member in $T$ is $>2$. Hence there does not
exist a $(k+t)$-robust equilibrium in the cheap talk game that can
$\epsilon$-implement the equilibrium with a mediator for any
$\epsilon<1$.
\end{proof}

\subsection{Proof of
\theoremref{thm:lowerbound-4}}\label{sec:proof-thm-4}

\rethm{thm:lowerbound-4}
If $2k + 3t < n \le 3k+3t$, there is a game $\Gamma$ and a
strong $(k,t)$-robust equilibrium $\sigma$ of a game $\Gamma_d$ with
a mediator $d$ that extends $\Gamma$ such that there exists a
$(k+t)$-punishment strategy with respect to $\sigma$ for which there
do not exist a natural number $c$ and a strategy  $\vecsigmact$ in the
cheap talk game extending $\Gamma$
such that the running time of $\vecsigmact$ on the equilibrium path is
at most $c$ and
$\vecsigmact$ is a $(k,t)$-robust implementation of $\sigma$.
\erethm

\begin{proof}
It remains to show that we can use $\sigma'$ as defined in the main text
to get a $(1,0)$-robust implementation in the 3-player mediator game
$\Gamma_{3,d}^{n,k+t}$, contradicting the argument above.
The idea is straightforward. Player $i$ in the 3-player game
simulates the players in $B_i$ in the $n$-player game, assuming that
player $j \in B_i$ has type $(j,f(j))$. (Recall that, in the
3-player game, player $i$'s type is a tuple consisting of $(j,f(j))$
for all $j \in B_i$.) In more detail, consider the strategy
$\sigma''_i$ where, in each round of the 3-player game, player $i$
sends player $j$ all the messages that a player in $B_i$ sent to a
player in $B_j$ in the $n$-player game (noting what the message is,
to whom it was sent, and who sent it).  After receiving a round $k$
message, each player $i$ in the 3-player game can simulate what the
players in $B_i$ do in round $k+1$ of the $n$-player game.
If a player $i$ does not send player $j$ a message of the right form in
the 3-player game, then all the
players in $B_j$ are viewed as having sent no message in the
simulation.  If all players in $B_i$ decide on the same value in the
simulation, then player $i$ decides on that value in the 3-player game;
otherwise, player $i$ decides 0.

It is easy to see that $\sigma''$ implements $\sigma$ in
$\Gamma_3^{n,k+t}$
and there is a bound $c$ such that all executions of $\sigma$ take
at most $c$ rounds, because $\vec{\sigma}'$ implements
$\vec{\sigma}^n$ in the $n$-player game and takes bounded time. It
follows from the argument above that $\vec{\sigma}''$ cannot be
(1,0)-robust.  Thus, some player $i$ must have a profitable
deviation. Suppose without loss of generality that it is player 3,
and 3's strategy when deviating is $\tau_3$.  Note that $\tau_3$ can
be viewed as prescribing what messages players in $B_3$ send to the
remaining players in the
game
$\Gamma^{n,k,t}_{CT}$.  (Recall, that if player 3 does not send
a message to player $j$ in the 3-player game that can be viewed as
part of such as description, and player $j$ is running $\sigma_j''$,
then player $j$ acts as if all the players in $B_3$ had sent the
players in $B_j$ no message at all. Thus, all messages from player
$i$ to player $j$ in the 3-player game can be interpreted as
messages from $B_i$ to $B_j$ in the $n$-player game.) Note that $t <
|B_3| \le k+t$.  (We must have $k \ge 1$, since otherwise we cannot
have $2k + 3t < n \le 3k+3t$, so $n \ge 3 +3t$.) Choose a subset $T$
of $B_3$ such that $|T| = t$. Let $\hat{\tau}_{B_3}$ be the strategy
in the $n$-player cheap-talk game whereby the players in $B_3$
simulate $\tau_3$, the players in $B_3 - T$ make the same decision
as player 3 makes using $\tau_3$, and the players in $T$ make the
opposite decision. It suffices to show that if all players in $B_3$
play $\hat{\tau}$, then the players in $B_3-T$ are better off than
they are playing $\sigma'$.

Note that every execution $r$ of $(\sigma'_{N-B_3},\hat{\tau}_{B_3})$ in the
cheap-talk extension of $\Gamma^{n,k,t}$ corresponds to a unique
execution $r'$ of $(\sigma''_{\{1,2\}},\tau_3)$ in the cheap-talk extension
$\Gamma_3^{n,k,t}$.  Thus, it suffices to show that the players in $B_3-T$
do at least as well in $r$ as player 3 does in $r'$.
Let $R_0'$, $R_1'$, and $R_2'$ be the set of executions of
$(\sigma''_{\{1,2\}}, \tau_3)$ where player 3 gets payoff $-3$, 1, and
2, respectively.  Let $R_j$ be the set of executions of
$(\sigma'_{N-B_3}, \tau'_{B_3})$ that correspond to an
execution of $R_j'$.  If $r' \in  R_0'$, then clearly a player in $B_3-T$
does at least as well in $r$ as player 3
does in $r'$.  If $r' \in R_1'$, then all three players play the secret
in $r'$.  Thus, in $r$, all the players in $B_3-T$ play the secret,
so they all get at least
1.
Finally, if $r' \in R_2'$, then in $r'$, player 3 plays the
secret, and either player 1 or 2 does not.
Hence
some player in $B_1$ or $B_2$ does not play the secret in the
cheap-talk extension of $\Gamma^{n,t,k}$. Moreover, all the players
in $T$ do not play the secret.  Thus, at least $t+1$ players do not
play the secret, so the players in $B_3 - T$ get 2. This completes
the argument.

\end{proof}

\subsection{Proof of \theoremref{thm:lowerbound-5a}}\label{sec:proof-thm-5a}\label{sec:thm-5a}

\rethm{thm:lowerbound-5a}
\lowerboundVa
\erethm

The proof uses a reduction to a generalization of the Byzantine
agreement problem called the $(k,t)$-\emph{\dagree problem},
which, as we said, is closely related to the problem of \emph{broadcast
with extended consistency} introduced by Fitzi~et al.~\citeyear{FHHW03}.
We have two parameters, $k$ and $t$. Each process has some initial
value, either $0$ or $1$. There are at most $k+t$ Byzantine
processes. Each non-Byzantine process must decide $0, 1,$ or
{\detect}. An execution is \emph{successful for Detect/Agree} if
the following three conditions all hold (the first two of which are
slight variants of the corresponding conditions in weak Byzantine
agreement):
\beginsmall{enumerate}

\item [I.] (Agreement:) If there are $t$ or fewer Byzantine processes,
then all non-Byzantine processes decide on the same value, and it is
a value in $\{0,1\}$.

\item [II.] (Nontriviality:) If all non-Byzantine processes have the
same initial value $v$ and no non-Byzantine process decides
{\detect}, then all the non-Byzantine processes must decide $v$.

\item[III.] (Detection validity:) If there are more than $t$ Byzantine
processes, then either all the non-Byzantine processes decide
{\detect}, or all non-Byzantine processes decide on the same value
in $\{0,1\}$.
\endsmall{enumerate}

Note that if $k=0$, then clause III is vacuous, so the
$(0,t)$-\dagree problem is equivalent to
Byzantine agreement with $t$ faulty processes \cite{LSP82}. Note
that the non-triviality condition for Byzantine agreement requires
all processes to decide $v$ if they all had initial value $v$, even
if there are some faulty processes.  Thus, it is a more stringent
requirement than the weak nontriviality condition of weak Byzantine
agreement.
The following argument, from which it follows that there does not
exist a protocol for the $(k,t)$-\dagree problem if $n \le 2k+3t$,
is based on a variant of the argument
used for \propref{lem:dist-lower-bound}.
If $T \subseteq N$, let $\vec{1}_T \vec{0}_{N-T}$ denote the input
vector where the players in $T$ get an input of 1 and the players in
$N-T$ get an input of 0.

\begin{proposition}\label{lem:dist-lower-bound1}
If
$\max\{2,t\} < n \le 2k+3t$ and $t \ge 1$, then for all
joint protocols $P$, there exist six scenarios $S_0, \ldots, S_5$,
six protocols $P^j_h$, for $j = 0,1$, $h=0,1,2$, and a partition of
the players into three
nonempty sets $B_0$, $B_1$, and $B_2$ such that $|B_0| \le t$,
$|B_1| \le k+t$, and $|B_2| \le k+t$, and
\beginsmall{itemize}
\item
in scenario $S_{0}$, the input vector is $\vec{0}$, in
$S_{1}$, it is $\vec{0}_{B_1 \cup B_2}\vec{1}_{B_0}$; and in
$S_{2}$, it is $\vec{0}_{B_2}\vec{1}_{B_0 \cup B_1}$;
in $S_{3+h}$, the input vector is the complement of the input vector
in $S_h$, for $h = 0, 1, 2$ (that is, if process $i$ has
input $\ell$ in $S_h$, it has input $1-\ell$ in $S_{3+h}$);
for all protocols $P^j$, for $j = 0,1$;
\item in $S_{3j+h}$ the processes in $B_h$ are faulty and use protocol
$P^j_h$, for $j= 0,1$ and $h=0,1,2$, while the remaining processes
are correct and use protocol $P$;
\item
the processes in $B_{h \oplus_3 2}$
receive exactly the same messages in every round of both
both $S_{3j + h}$ and $S_{3j + h \oplus_6 1}$
(where we use $\oplus_\ell$ and $\ominus_\ell$ to denote addition
and subtraction mod $\ell$).
\endsmall{itemize}
\end{proposition}

\begin{proof}
\begin{figure}[tb]
  \centerline{\includegraphics[height=4in, width=4in, scale=4]{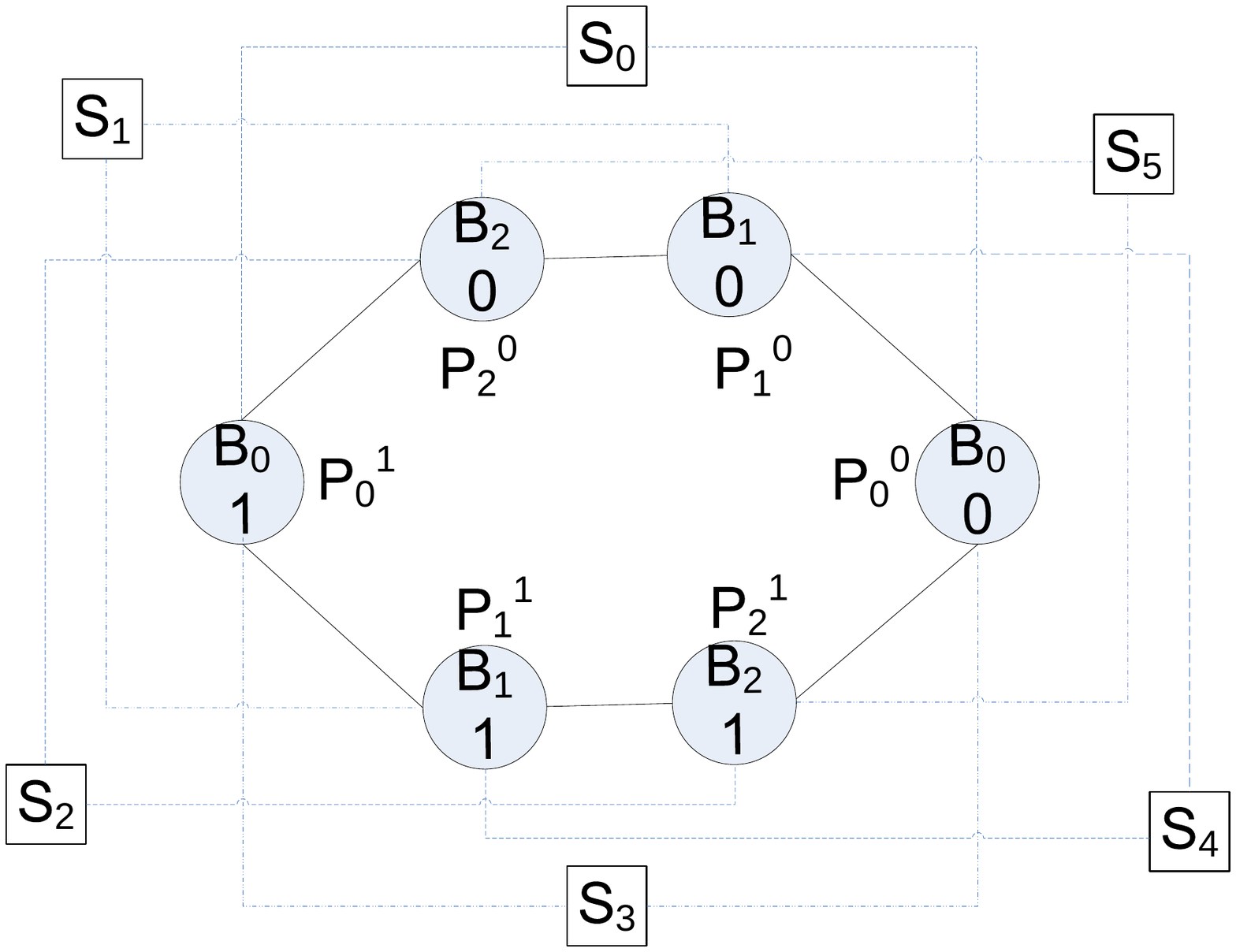}}
  \vskip-78pt
\caption{The construction for the proof of
\propref{lem:dist-lower-bound1}.} \vskip-8pt \label{fig:hexagon}
\end{figure}

We explicitly construct the scenarios and protocols. Given $P$, we
simultaneously describe the protocols $P^j_h$ and scenarios
$S_{3j+h}$, for $j=0,1$, $h=0,1,2$, by induction, round by round.
Suppose we have defined the behavior of protocol $P^j_h$ and the
scenario $S_{3j+h}$ for the first
$\ell$ rounds. As required by the proposition, in scenario
$S_{3j+h}$, the processes in
$B_h$ are faulty and use protocol $P_h^j$, the correct processes use
protocol $P$, and the inputs are as required by the protocol. The
hexagon in \figureref{fig:hexagon}
implicitly defines the scenarios
and what happens in
round $(\ell+1)$. In round $\ell+1$ of scenario $S_{3j + h}$, the
processes in $B_h$ are faulty; each process $i \in B_h$ sends
messages to the processes in $B_{h \oplus_3 1}$ as if $i$ is using
$P_i$ and in the first $k$ rounds has received exactly  the messages
it would have received in scenario $S_{3j + h \ominus_6 1}$, and
sends messages to the processes in $B_{3j + h \ominus_3 1}$ as if
$i$ is using $P_i$ and has received exactly the same messages it
would have received in the first $k$ rounds of $S_{3j + h  \oplus_6
1}$. Note that in scenarios $S_{3j + h \oplus_6 1}$ and $S_{3j + h
\ominus_6 - 1}$,
a process $i \in B_h$ uses protocol $P_i$.  Thus, $i \in B_h$ sends
the same message to processes in $B_{h\oplus_3 1}$ in both scenarios
$S_{3j + h}$ and $S_{3j + h \oplus_6 1}$.  Finally, note that there
is no need for processes in $B_h$ to send
messages to other processes
in $B_h$ in scenario $S_{3j+h}$. The behavior of the processes in
$B_h$ does not depend on the messages they actually receive in
scenario $S_{3j+h}$; they are simulating scenarios $S_{3j+h \oplus_6
1}$ and $S_{3j+h \ominus_6 1}$. This behavior characterizes protocol
$P_j^h$.

The six scenarios and protocols are implicitly defined by the
hexagon
in \figureref{fig:hexagon}.
For example, scenario $S_0$ is defined by the four nodes in the
hexagon starting with the one labeled $P_0^0$ and going
counterclockwise. The inputs for the processes in $B_h$ are defined
by the numbers in circle in the first three nodes; in this scenario,
all processes have input 0, because these numbers are all 0.   The
processes in $B_0$, which are the faulty ones in this scenario,
behave to the processes in $B_1$ as if they have input 0 and to the
processes in $B_2$ as if they have input 1.  (This is indicated by
the edge joining the node labeled $B_0 (0)$ and $B_1 (0)$ and the
edge joining $B_0 (1)$ and $B_2 (0)$).

\commentout{
\begin{verbatim}

      B_1 (0)
     /   \
   /       \
B_2 (0)   B_3 (1)
  |         |
  |         |
B_3 (0)   B_2 (1)
   \       /
     \   /
     B_1 (1)

\end{verbatim}
} %

By construction, the processes in $B_{h \oplus_3 2}$ are correct in
both $S_{3j + h}$ and $S_{3j + h \oplus_6 1}$; an easy proof by
induction on rounds shows that they receive exactly the same
messages in every round of both scenarios.
\end{proof}

The following is a generalization of the well-known result that
Byzantine agreement (which, as we have observed, is just the
$(0,t)$-\dagree problem) cannot be solved if $n \le 3t$.

\begin{proposition}\label{prop:dagree}
If $n \le 2k+3t$ and $t \ge 1$, then there is no protocol that
solves the $(k,t)$-\dagree problem.
Moreover, if there is a partition of the players into three nonempty
sets $B_0$, $B_1$, and $B_2$, as in
\propref{lem:dist-lower-bound1},
and
the four input vectors $\vec{0}$, $\vec{1}_{B_0}\vec{0}_{B_1 \cup
B_2}$, $\vec{1}_{B_0 \cup B_1} \vec{0}_{B_2}$, and $\vec{1}$ each
have probability $1/4$ then, for any protocol $P$, there exists a
set $T$ with $|T| \le k+t$
and protocol  $P'$ such that the  probability that an execution of
$(P_{N-T},P'_T)$ is unsuccessful for $(k,t)$-Detect/Agree is at
least $1/20$.
\end{proposition}

\begin{proof}
Suppose that $n \le 2k + 3t$ and that $P$ solves the $(k,t)$-\dagree
problem.  Consider the six scenarios from
\propref{lem:dist-lower-bound1}.  Since $t \ge 1$, each of
$B_0$, $B_1$, and $B_2$
is nonempty.  In scenario $S_0$, all the
correct processes must decide on 0, by the nontriviality condition.
By \propref{lem:dist-lower-bound1},
the processes in $B_2$ cannot distinguish $S_0$ from $S_1$ and
are correct in both,
therefore the correct processes
must decide 0 in $S_1$.  Thus, by the
agreement property (if
$|B_0| \le t$)
or the detection validity
property
(if $|B_2| > t$),
all the processes in
$B_1$
must also
decide $0$ in $S_1$.  A similar argument shows that all the
processes in $B_0 \cup B_1$ must decide 0 in $S_2$ and all the
processes in $B_1 \cup B_2$ must decide 0 in $S_3$.  But in $S_3$
the nonfaulty processes must decide 1.
This proves the first part of the claim.

Now
suppose that each of $\vec{0}$, $\vec{1}_{B_0}\vec{0}_{B_1 \cup
B_2}$, $\vec{1}_{B_0 \cup B_1} \vec{0}_{B_2}$, and $\vec{1}$ have
probability $1/4$.  Again, consider the scenarios $S_0, \ldots,
S_3$.
We claim that, for some $j \in \{0,1,2,3\}$, if the faulty processes
use the
strategies prescribed for scenario $S_j$, then conditional on the
input vector being that of scenario $S_j$, the probability that an
execution is unsuccessful for $(k,t)$-Detect/Agree is at
least $1/5$. We prove by induction on $j$ that either the claim
holds for some scenario $S_j$ with $0 \le j \le 3$ or the
probability that each correct process in scenario $S_j$ decides 0 is
greater than $1 - (j+1)/5$. For the base step, suppose that in
scenario $S_0$ some correct process decides 1 with probability
greater than $1/5$.  Then if processes in $B_0$ are faulty and use
protocol $P_0^0$, no matter what their input, if the input vector is
actually $\vec{0}$
the execution will be unsuccessful with probability greater than
$1/5$.
Assume now that the claim does not hold for processes in $S_0$.
Since processes in $B_2$ are correct in both $S_0$ and $S_1$, and
cannot distinguish $S_0$ from $S_1$, they must decide 0 with
probability at least $4/5$ in $S_1$.  A similar argument now shows
that either the claim holds for $S_1$, or the processes in $B_0$
must decide 0 with probability at least $3/5$ in $S_1$.  The
inductive step is similar, and left to the reader.  We complete the
proof by observing that in $S_3$, if the claim does  not hold, then
all the processes must decide 0 with probability at least $1/5$.
But in $S_3$, all processes are correct and have initial value 1.
Thus, again the probability of an unsuccessful execution is at least
$1/5$.  Since each relevant input vector has probability $1/4$, the
probability of an unsuccessful execution is at least $1/20$.
\end{proof}

\begin{proof} [Proof of \theoremref{thm:lowerbound-5a}]
Suppose that $2k + 2t < n \le 2k + 3t$, $t\ge 1$. Consider the same
game and strategies as in the proof of \theoremref{thm:lowerbound-2}
with utility vector $u^M$, where $M
> 20(1 + \epsilon)$.
By \lemmaref{lem:lowerbound-2}, $\sigma$ is a $(k,t$)-robust
strategy in the mediator game $\Gamma_d$.
Suppose,  by way of contradiction, that $\sigma'$ is an \eps
$(k,t)$-robust  implementation of  $\sigma$ in the CT extension
$\Gammact$ of $\Gamma_{d}(u^M)$.  Let $P_i$ be the protocol for
process $i$ that sends the same messages and makes the same
decisions as $\sigma'_i$.
By \propref{prop:dagree}, there is a protocol $P'$ and a set $T$
with $|T| \le k+t$ such that the probability that an execution of
$(P_{N-T},P'_T)$ is unsuccessful for $(k,t)$-Detect/Agree is at
least $1/20$.   Let $\sigma''$ be the strategy where the players
choose characteristic $B$ and play according $P'$ in the cheap-talk
game,  no matter what their actual input. With probability $1/20$,
the outcome will be bad (since an unsuccessful outcome with
$(P_{N-T},P'_T)$ corresponds to a bad outcome with
$(\sigma'_{N-T},\sigma''_T)$. Thus, the expected utility for the
players in $T$ is greater than $1 + \epsilon$, so $\sigma'$ is not
\eps $(k,t)$-robust.
\end{proof}

\subsection{Proof of \theoremref{pro:lowerboundbii}}\label{sec:-proof-pro:lowerboundbii}\label{thm6proof}

To prove \theoremref{pro:lowerboundbii}, we start with the case that
$n=2$, $k=1$, and $t=0$. The ideas for this proof actually go back
to Shamir, Rivest, and Adleman \cite{SRA81}; a similar result is
also proved by Heller \cite{Heller05}.  However, these earlier
proofs assume that in the cheap-talk protocol, the players first
exchange messages and then, after the message exchange, make their
decision (perhaps using some randomization).  That is, they are
implicitly assuming that when the cheap-talk phase of the strategy
has ended, it is common knowledge that it has ended (although when
it ends may depend on some random choices).  While this is a
reasonable assumption if we have a bounded cheap-talk protocol, our
possibility results involve cheap-talk games with no a priori upper
bound on running time.  We do not want to assume that the players
receive a signal of some sort to indicate that the message exchange
portion of the cheap-talk has ended.  Our lower bound proof does not
make this assumption.
We can, of course, find a round such $b$
that with high probability.  This solves part of the problem.  However,
the earlier proofs also took advantage of the fact that players decide
\emph{simultaneously}, after the cheap-talk phase ends.  Since we do not
make this assumption, our proof requires somewhat more delicate
techniques than the proofs in the earlier papers.

\begin{proposition}\label{pro:Heller}
If $n=2$, then there exist a game $\Gamma$, $\epsilon > 0$, a
mediator game $\Gamma_d$ extending $\Gamma$, a Nash equilibrium
$\sigma$ of $\Gamma_d$, and a punishment strategy $\rho$ with
respect to $\sigma$ such that there is no strategy $\sigma'$ that is
an
$\eps (1,0)$-robust implementation of $\sigma$.
\end{proposition}

\begin{figure}[tb]
\center{
\begin{minipage}[c]{2.5in}{
\begin{tabular}{l|c|c|}\multicolumn{1}{l}{}&\multicolumn{1}{c}{$L$}&\multicolumn{1}{c}{$R$}\\
\cline{2-3}
 $U$& $(3,3)$& $(1,4)$ \\ \cline{2-3}
 $D$&$(4,1)$&$(0,0)$\\ \cline{2-3}
\multicolumn{1}{l}{\vspace{-3mm}}
&\multicolumn{1}{c}{\ \hspace{15mm}\ }&\multicolumn{1}{c}{\ \hspace{15mm}\ }\\
\multicolumn{1}{l}{}&\multicolumn{2}{c}{A simple 2-player game.}\\
\multicolumn{3}{l}{}\\
\end{tabular}\\
}
\end{minipage}
\
\begin{minipage}[htb]{2.5in}
\begin{tabular}{l|c|c|}\multicolumn{1}{l}{}&\multicolumn{1}{c}{$L$}&\multicolumn{1}{c}{$R$}\\
\cline{2-3}
 $U$& $1/3$& $1/3$ \\ \cline{2-3}
 $D$&$1/3$&$0$\\ \cline{2-3}
\multicolumn{1}{l}{\vspace{-3mm}}
&\multicolumn{1}{c}{\ \hspace{15mm}\ }&\multicolumn{1}{c}{\ \hspace{15mm}\ }\\
\multicolumn{1}{l}{}&\multicolumn{2}{c}{A correlated equilibrium.}
\\
\multicolumn{3}{l}{}\\
\end{tabular}\\
\end{minipage}
\vskip-18pt
\caption{The game used in the proof of  \propref{pro:Heller}}
\vskip-8pt
\label{fig:tables}}
\end{figure}

\begin{proof}
Let $\Gamma$ be the game described in the left table of
\figureref{fig:tables},
where
player 1 is Alice and player 2 is Bob, Alice can choose between
actions $U$ and $D$, and Bob can choose between $L$ and $R$. The
players all have a single type in this game, so we do not describe
the types.  The boxes in the left table describe the utilities of
Alice and Bob for each action profile.  The right table describes a
correlated equilibrium of this game, giving the probabilities that
each action profile is played.

Consider the mediator game $\Gamma_d$ extending $\Gamma,$ where the mediator
recommends the correlated equilibrium described in
the table on the right of \figureref{fig:tables}
(that is, the mediator recommends
choosing an  action profile with the probability described in the table,
and recommends that each player play his/her component of the
action profile).  Let $\sigma$ be the strategy profile of following the
mediator's recommendation. It is easy to see that $\sigma$ is a Nash
equilibrium of the mediator game; moreover, $(D,R)$ is a punishment
strategy with respect to $\sigma$.  Also, note that the requirement
of a broadcast channel trivially holds if $n=2$.

Suppose, by way of contradiction, that
$\sigma'=(\sigma_1',\sigma_2')$ is a $(1/10)$-Nash equilibrium that
implements
$\sigma$ in a CT extension $\Gammact$ of $\Gamma_{d}$.  Since
$\sigma'$ implements $\sigma$, it must be the case that, with
probability 1, an execution of $\sigma'$ terminates. Hence, there
must be some round $b$ such that, with probability at least
$1-\frac{1}{\beta^2},$ $\sigma'$ has terminated (with both players
choosing an action in the underlying game) by the end of round $b$.
(We determine $\beta$ shortly.)

The execution of $\sigma'$ is completely determined by the random
choices made by the players.  Let $\R_i$ denote the set of possible
sequences of random choices by player $i$.  (For example, if player
$i$ randomizes by tossing coins, then $r \in \R_i$ can be taken to
be a countable sequence of coin tosses.) For ease of exposition, we
assume that $i$ makes a random choice at every time step (if the
move at some time step in the cheap-talk game is deterministic, then
$i$ can ignore the random choice at that time step).  If $r$ is a
sequence of random choices, we use $r^\ell$ to denote the
subsequence of $r$ consisting of the random choices made in the
first $\ell$ steps. Since $\sigma'$ implements $\sigma$, with
probability 1, both players choose an action in the underlying game
in an execution of $\sigma'$. (That is, while it is possible that
there are infinite executions of $\sigma'$ where some party does not
choose an action, they occur with probability 0.) Let $\R = \R_1
\times \R_2$.  Note that the probability on the random sequences in
$\R$ determines the probability of outcomes according to $\sigma'$.

Suppose that $r$ is a finite random sequence of length $\ell$ for
player $i$.
We take $\Pr(r) = \Pr(\{s \in \R_i: s^\ell = r\})$.  We
similarly define $\Pr(r_1,r_2)$ for a pair $(r_1,r_2)$ of finite
sequences of equal length. The random sequences determine the
message history and the actions. Given random sequences $r_1$ and
$r_2$ of length $\ell$, let $H(r_1,r_2)$ be the pair of message
histories $(h_1,h_2)$ determined by $(r_1,r_2)$, and let $A(r_1,r_2)
= (a_1,a_2)$ be the action profile chosen as a result of $(r_1,r_2)$
(where we take $a_i$ to be $\bot$ if player $i$ has not yet
taken an action); in this case, we write $A_i(r_1,r_2) = a_i$, for $i = 1,
2$.

A pair of histories $(h_1,h_2)$ of equal length is
\emph{deterministic for player $i$} if it is not the case that both
of player $i$'s actions have positive probability, conditional on
the message history being $(h_1,h_2)$. We claim that all history
pairs that arise with positive probability must be deterministic
for some player $i$.
For suppose that a history pair $(h_1,h_2)$ is not deterministic for
either player.
Let $\R_1' = \{(r_1, r_2) : A_1(r_1,r_2) = D, H(r_1, r_2) = (h_1,
h_2)\}$ and let $\R_2' = \{(r_1, r_2) : A_2(r_1,r_2) = R, H(r_1,
r_2) = (h_1, h_2)\}$.  By assumption, $\Pr(\R_1' \mid (h_1,h_2)) >
0$ and $\Pr(\R_2' \mid (h_1,h_2)) > 0$.  Now let $\R_3' =
\{(r_1,r_2): \exists r_1', r_2' ((r_1,r_2') \in
\R_1', (r_1',r_2) \in \R_2'\}$.
It is easy to see that for $(r_1,r_2) \in \R_3'$, we have
$H(r_1,r_2) = (h_1,h_2)$ (we can prove by a straightforward
induction that $h(r_1^\ell,r_2^\ell) = (h_1^\ell, h_2^\ell)$ for
each round $\ell$ less than or equal to the length of $r_1$). Hence
$A(r_1,r_2) = (D,R)$.  Moreover, $\Pr(\R_3' \mid (h_1,h_2)) \ge
\Pr(\R_1'\mid (h_1,h_2)) \times \Pr(\R_2'\mid (h_1,h_2)) > 0$. Thus,
if $(h_1,h_2)$ has positive probability, then the outcome $(D,R)$
has positive probability, which contradicts the assumption that
$\sigma'$ implements $\sigma$.

Consider the following two strategies $\sigma_1''$ and $\sigma_2''$
for players 1 and 2, respectively.  According to $\sigma_1''$,
player 1 sends exactly the messages that he would have sent
according to $\sigma_1'$ until round $b$, but does not take an action until
the end of round $b$.  If player 1 observes the histories
$(h_1,h_2)$ at the end of $b$ rounds, and
if
the probability of player
2 playing
$L$
conditional on having observed $(h_1,h_2)$ is at least
$1-1/\beta,$
then player 1 decides $D$, but keeps sending
messages according to $\sigma_1'$;
otherwise, player 1 plays exactly according to $\sigma_1'$.
(Note that computing whether to play $U$ may be difficult, but we are
assuming computationally unbounded players here.)
The strategy $\sigma_2''$ is
similar, except now player 2 will play $R$ if conditional on
$(h_1,h_2)$, player 1 plays $U$ with probability at least
$1-1/\beta.$
It is easy to see that if player $1$ plays a different action with
$\sigma_1''$ when observing $(h_1,h_2)$ than with $\sigma_1'$, it must be
because $\sigma_1'$ recommends $U$ and player 1 plays $D$. Moreover,
player 1's expected gain in this case,
conditional on observing
$(h_1,h_2)$
(given that player 2 plays $\sigma_2'$)
is at least
$1\cdot(1-1/\beta) -3\cdot(1/\beta)=1-4/\beta.$
Similarly, if player 2 plays a
different action when observing $(h_1,h_2)$, then player 2's
expected gain conditional on observing $(h_1,h_2)$
and that player 1 plays $\sigma_1'$
is at least
$1-4/\beta.$

Let $\R'(U,L) = \{(r_1,r_2): A(r_1,r_2) = (U,L),$ $r_1$, $r_2$ have
length $b\}$. Since $\sigma'$  implements $\sigma$, the probability
of the outcome $(U,L)$ with $\sigma'$ must be $1/3$. Since $b$ was
chosen such that the probability of not terminating within $b$
rounds is less than $1/\beta^2,$ we must have $\Pr(\R'(U,L)) > 1/3 -
1/\beta^2.$

Let $H'$ consists of all message histories $(h_1,h_2)$ of length $b$
such that, with probability at least $1/\beta$, at least one player
does not terminate by the end of round $b$ (where the probability is
taken over pairs $(r_1,r_2)$ such that $H(r_1,r_2) = (h_1,h_2)$).
The probability of $H'$ must be at most $1/\beta$ (otherwise the
probability of not terminating by the end of round $b$ would be
greater than $1/\beta^2).$ Thus, for any history that is not in
$H'$, the probability that both players take an action in $\sigma'$
is at least $1-1/\beta.$

Let $\R''(U,L) = \{(r_1,r_2) \in \R'(U,L): H(r_1,r_2)\not\in H'\}.$
The discussion above implies
that
the inequality $\Pr(\R''(U,L)) > 1/3 -
1/\beta^2-1/\beta$
holds.
By the arguments above,
at least half of the histories in
$\R''(U,L)$ are deterministic for one of the players.
Without loss of
generality, let it be player 1
(Alice).
With probability at least $1/2\cdot(1/3
- 1/\beta^2 - 1/\beta)$,
Alice
would have made a choice of
$U$ by the end of round $b$, if she had taken an action. With
probability $1-1/\beta$ she will take an action, and therefore Bob
can play $\sigma_2''$ and will gain an expected utility
$(1-4/\beta)\cdot 1/2\cdot(1/3 - 1/\beta^2 - 1/\beta).$   We can
choose $\beta$ such that the expected gain is at least $1/10$. Thus,
$\sigma'$ is not a $(1/10)$-equilibrium.
By multiplying all utilities in $\Gamma$ by $10\epsilon$, we get a game
$\Gamma^\epsilon$ such that $\Gamma^\epsilon_d$ has a (1,0)-robust
equilibrium that has no $\epsilon$-implementation.
\end{proof}

We now prove \theoremref{pro:lowerboundbii} by generalizing this
result to arbitrary $k$ and $t$.

\rethm{pro:lowerboundbii}
If $k + 2t < n \leq 2(k+t)$
there exist a game $\Gamma$,
an
$\epsilon > 0$,
a mediator
game $\Gamma_d$ extending $\Gamma$, a
strong
$(k,t)$-robust
equilibrium $\sigma$ of $\Gamma_d$, and a
$(k+t)$-punishment strategy $\rho$ with respect to $\sigma$ such
that there is no strategy $\vecsigmact$ that is
an
\eps
$(k,t)$-robust implementation of $\sigma$ in the cheap-talk
extension of $\Gamma$, even with broadcast channels. \erethm

\begin{proof}
Divide the players into three
disjoint
groups: group $A_1$ and $A_2$ have
each have $n-(k+t)$ members, and group $B$ has $2k + 2t - n$
members.  It is immediate that this can be done, since $2(n-(k-t)) +
2k + 2t - n = n$. Moreover, $|A_1 \cup B| = |A_2 \cup B| = k+t$.
Note that neither $A_1$ nor $A_2$ is empty; however, $B$ may not
have any members.

Players do not get any input (i.e., there is only one type).
Intuitively, a player must output a value in field $F$, with $|F| \ge 6$,
signed by a check
vector.  More precisely,
a player $i \in A_1 \cup A_2$ outputs  $8(n-1) + 2$ elements of $F$, and
optionally
either $U$ or $D$ if $i \in A_1$ or $L$ or $R$ if $i \in A_2$.
The first two elements of $F$ output by $i$ are denoted $a_{i1}$ and
$a_{i2}$.  We think of these as $i$'s share of two different secrets.
The remaining $8(n-1)$ elements of $F$ consist of $n-1$ tuples of 8
elements, denoted $(y_{1ij},y_{2ij}, b_{1ji}, b_{2ji},
b_{3jij}, c_{1ji},c_{2ji}, c_{3ji})$.
We have one such tuple for each player $j \ne i$.  We require that
neither $b_{1ji}$ nor $b_{2ji}$ is 0.
A player $i$ in group $B$ must
output $3 + 9(n-1)$ numbers, denoted
$a_{hi}, y_{hij}, b_{hji}, c_{hji}$, for $h = 1, 2, 3$ and
for each player $j \ne i$ (again, we require that $b_{hji} \ne 0$);
together with an optional
element, which can be any of $U$, $D$, $L$, or $R$.
For each player $j \ne i$, we would like to have
\begin{equation}\label{eq:reliable}
a_{hi}+y_{hij}b_{hij}=c_{hij},
\end{equation}
for $h = 1,2$; if $i \in B$, then we
also would like to have $a_{3i}+y_{3ij}b_{3ij}=c_{3ij}$.
Note that the field elements $a_{hi}$ and $y_{hij}$ are output by player
$i$, while $b_{hij}$ and $c_{hij}$ are output by player $j$.
As we said, we think of the values $a_{hi}$ to be shares of some
secret.  Thus, we would like there to be a function $f_1$
which interpolates the values $a_{1i}$; $f_1(0)$ is intended to encode
an action in $\{U,D\}$ for the players in $A_1$ to play.
That is, if $f_1(0) = 0$, then the players in $A_1$ should play $U$; if
$f_1(0)
= 1$, then they should play $D$.)
Similarly we
would like there to be a function $f_2$ that encodes an action in
$\{L,R\}$ for the players in $A_2$ to play.  Finally, the values $a_{3i}$
should encode a pair of values in $\{U,D\} \times \{L,R\}$, where
$(U,L)$ is encoded by 0, $(U,R)$ by 1, $(D,L)$ by 2, and $(D,R)$ by 3.

The payoffs are determined as follows, where we take a player to be
\emph{reliable} if
there exists a set $N'$ of $n-1-t$ agents $j \ne i$
such that Eq.~(\ref{eq:reliable})
holds for $k=1$ and $k=2$, and for $k=3$
if $i \in B$.
\begin{itemize}
\item If more than $t$ players are unreliable, then all players get 0.

\item If all the players in group $A_i \cup B$ play the same optional value and some are
unreliable, then all players get 0.
\item If $t$ or fewer players are unreliable, but
either (a) there does not exist a unique polynomial $f_1 $ of degree $k+t-1$
that interpolates
the values $a_{1i}$ sent by the reliable players in $i \in A_1 \cup B$
such that $f_1(0) \in \{0,1\}$;
(b) there does not exist a unique polynomial $f_2 $ of degree $k+t-1$
that interpolates
the values $a_{1i}$ sent by reliable players $i$ in group $A_2$ and
the values $a_{2i}$
sent by reliable players in group $B$
such that $f_2(i) \in \{0,1\}$;
(c) $(f_1(0), f_2(0))$ encodes $(D,R)$;
or (d) there does not exist
a unique polynomial $f_3$ of degree $k+t-1$ that interpolates the
values $a_{2i}$ sent by reliable players $i \in A_1 \cup A_2$ and
the  values $a_{3i}$ sent by reliable players $i \in B$
such that $f_3(0)$ encodes $(f_1(0),f_2(0))$, in the sense described
above,
then all players get $8/3$;

\item if (a), (b), (c), and (d) above all
do not
hold, then suppose that $g(0)$ encodes $(x,y)$.
Let $o_1$ be $x$ unless everyone in $A_1 \cup B$
plays $x'$ as their optional value, where $x' \in \{U,D\}$, in
which case $o_1 = x'$. Similarly, let $o_2$ be $y$ unless everyone in
groups $A_2 \cup B$ plays $y'$ as their optional value,
where $y \in \{L,R\}$,  in which case $y' = o_2$.  Let $(p_1,p_2)$ be
the payoff according to $(o_1,o_2)$, as described in
\figureref{fig:tables}.
Then everyone in group $A_i$ gets $p_i$, for $i = 1,2$.
Payoffs for players in $B$ are determined as follows: if
everyone in $A_1 \cup B$ played $x' \in \{U,D\}$ as their
optional value, then players in $B$ gets $p_1$; if everyone
in $A_2 \cup B$
played
$y' \in \{L,R\}$ as their optional value, then
everyone in $B$ gets $p_2$; otherwise, everyone in $B$ gets $8/3$.
\end{itemize}

The mediator chooses an output $(o_1,o_2)$ according to the
distribution described in
\figureref{fig:tables}.
The mediator then
encodes $o_i$ as the secret of a degree $k+t-1$ polynomial $f_i$;
that is, $o_i = f_i(0)$ and encodes $(o_1,o_2)$ as the secret of a
degree $k+t$ polynomial $g$. Suppose players $1, \ldots n-(k+t)$ are
in group $A_1$; players $n-(k+t) + 1, \ldots, 2(n-(k+t))$ are in
group $A_2$; and players $2(n-(k+t)) + 1, \ldots, n$ are in group
$B$. The mediator sends each player $j$ in group $A_i$ $(f_i(j),
g(j))$, for $i = 1, 2$, and sends each player $j$ in group $B$
$(f_1(j),f_2(j),g(j))$.
In addition, the mediator sends all players consistent check vectors
such that $a_1(i)+y_{1ij}b_{1ij}=c_{1ij}$,
$a_2(i)+y_{2ij}b_{2ij}=c_{2ij}$ and $a_3(i)+y_{3ij}b_{3ij}=c_{3ij}$
for all $i,j \in N$.
If the players play the message sent by the
mediator (and do not play the optional value), then they get
expected payoff $8/3$.

We now show that this strategy is $(k,t)$ robust.  For $t$-immunity,
note that the $t$ players cannot take over all of $A_i \cup B$ for $i
= 1$ or $i = 2$ (since both of these sets have cardinality $k+t$), so they
cannot take advantage of sending the optional element of $\{U,D,L,R\}$.
If any of the $t$ players are shown to be unreliable, it is easy
to see that this cannot hurt the other players, since there will not be
more than $t$ unreliable players.  If the reliable players do not send
values that pass checks (a), (b), (c), and (d) above, then each player
gets  $8/3$, which is the expected payoff of playing the recommended
strategy. Finally, if a large enough subset of the $t$ players manage to
guess the check vectors and send values that satisfy (a), (b), (c), and
(d) above, because they do not know any of the secrets, they are
effectively making a random change to the output, so the expected payoff
is still $8/3$.

For $(k,t)$ robustness, note that a set of $k+t$ that consist of all of
$A_i \cup B$ for $i = 1$ or $i = 2$ can change the output by all
playing the same optional value, but they cannot improve their payoff
this way, since we have a correlated equilibrium in the 2-player game.
If all $k+t$ players are in group $A_1 \cup B$, and their value is
$U$, they can try to change the outcome to $(D,L)$ by all playing $D$
and guessing shares and check values
in the hope
of changing them to
the $(D,L)$ outcome.  But if they are caught, they will all get 0.
Since $|F| \ge 6$, the probability of getting caught is greater than $1/4$,
so they do not gain by deviating in this way.

Note that if $n > k + 2t$, then if $n-(k+t)$ players deliberately do
not play the values sent them by the mediator, then this is a
$(k+t)$-punishment strategy with respect to $\sigma$, since $n -
(k+t) > t$.

Finally, the same argument as in the 2-player game shows that, by taking
over either $A_1 \cup B$ or $A_2 \cup B$, a set of size $k+t$ can
improve their outcome by deviating.
\end{proof}

\subsection{Proof of
\theoremref{thm:d-broadcast-lowerbound}}\label{sec:proof-d-broadcast-lowerbound}

\rethm{thm:d-broadcast-lowerbound}
\lowerboundX
\erethm

\begin{proof}
We consider
a relaxation of Byzantine agreement that we call the
\emph{$(k,t)$-partial broadcast problem}. There are $n$ processes
and process $1$ is designated as leader. The leader has an initial
value 0 or 1. Each process must decide on a value in $\{0, 1,
\pass\}$. An execution of a protocol $P$ is \emph{successful for
$(k,t)$-partial broadcast} if the following two conditions hold:

\beginsmall{enumerate}
\item [I.] (Agreement): If there are $t$ or fewer Byzantine processes
and the leader is non-Byzantine then all non-Byzantine processes
decide on the leader's value.

\item [II.] (No disagreement): If there are $k+t$ or fewer Byzantine processes,
then there do not exist two non-Byzantine processes such that one
decides 0 and the other decides 1.
\endsmall{enumerate}

Note that if the leader is faulty, it is acceptable that some
non-Byzantine processes decide on a common value $v \in \{0,1\}$ and
all other non-Byzantine processes decide $\pass$. Observe that the
$(0,t)$-partial broadcast problem is a relaxation of the well-known
\emph{Byzantine generals} problem \cite{LSP82}. We provide
probabilistic lower bounds for
this problem, which also imply known
probabilistic lower bounds for the Byzantine generals problem.

\begin{proposition}\label{prop:Byzgen}
If $\max(2,k+t) < n \le k + 3t$ and each input for the leader is
equally likely.
Then there exists a function $\Psi$ that maps protocols to protocols
such that for all joint protocols $P$, there exists a set $T$ of
processes such that either
\begin{enumerate}
\item[(a)] $|T|\leq t$, $1 \notin T$ and the execution $(P_{N-T},
\Psi(P)_T)$ fails the agreement property with probability at least 1/6;
or
\item[(b)] $|T|\leq k+t$, $1\in T$ and the execution $(P_{N-T},
\Psi(P)_T)$ fails the no-disagreement property with probability at least
1/6.
\end{enumerate}
The running time of $\Psi(P)$ is polynomial in the number of players
and the running time of $P$.
\end{proposition}

\begin{proof}
Partition the $n$ players into 3 nonempty sets $B_0$, $B_1$, and
$B_2$ such that $|B_0|\le t$, $|B_1|\le k+t$, and $|B_2|\le t$.
Assume that $B_1$ contains process 1 (the leader). Consider the
scenario consisting of $2n$ processes partitioned into six sets
$A_0,A_1,\dots,A_5$ such that the processes of $A_i$ have the
indexes of the processes of $B_{i \pmod 3}$; messages sent by
processes in $A_i$ according to $P$ reach the appropriate recipients
in $A_{i-1 \pmod 6},A_i,A_{i+1 \pmod 6}$. For example, if a
processes $\ell \in A_i$ has index $j \in N$ and is supposed to send
a messages to a processes with index $j' \in N$ such that $j' \in
B_{i' \pmod 3}$ and $i' \in \{i-1,i,i+1\}$ then this message will
reach the process $\ell' \in A_{i'}$ whose index is $j'$; the
process with index 1
(the leader) starts with initial value 1 in $A_1$ with
with initial value 0 in $A_4$.

Any two consecutive sets $A_i,A_{i +1 \pmod 6}$ define a possible
scenario denoted $S_i$, where the non-faulty processes, $A_i \cup
A_{i + 1 \pmod 6}$, execute $P$, and the faulty processes simulate
the execution of all the 4 remaining sets of processes. For $i \in
\{0,1,3,4\}$, let $e_i$ denote the probability that protocol $P$
fails the agreement condition in scenario $S_i$; for $i \in \{2,5\}$
let $e_i$ denote the probability that protocol $P$ fails the
no-disagreement condition in scenario $S_i$.

We claim that $e_2 \geq 1-(e_1 +e_3)$. Indeed with probability
$1-e_1$ processes in $A_2$ succeed in $S_1$ which implies that
processes in $A_2$ must decide 1 in this case. Similarly, with
probability $1-e_3$ processes in $A_3$ must decide 0 due to $S_3$,
hence in $S_2$, the non-faulty processes must reach disagreement
with probability at least $1-(e_1+e_3)$. A symmetric argument gives
$e_5 \geq 1-(e_4+e_0)$.

Therefore it cannot be the case that for all $i\in\{0,1,2\}$,
$e_i+e_{i+3} < 2/3$. Let $i$ be an index such that $e_i+e_{i+3}
\geq 2/3$ and consider the set
$B_{i - 1 \pmod 3}$
of processes that are
Byzantine and simulate 4 sets of processers according to scenario
$S_i$ or $S_{i+3 \pmod 6}$ with uniform probability.

Given that $e_i+e_{i+3 \pmod 6} \geq 2/3$ then the expected
probability of failure is $1/3$ if the faulty players know the
initial value. Since they can guess the initial value, the faulty
players in $B_{i-1 \pmod 3}$ will cause $P$ to fail with probability
at least $1/6$. If $i=1$ then $|B_i|\leq t+k$, $1 \in B_i$ and the
no-disagreement condition fails, and if $i \in \{0,2\}$ then
$|B_i|\leq t$, $1 \notin B_i$ and the agreement condition fails.
\end{proof}

We now construct a game that captures the $(k,t)$-partial broadcast
problem.  Given $k$ and $t$, consider the following game $\Gamma$
with $n$ players. Player 1 is the \emph{broadcaster} and has two
possible types, 0 or 1, both equally likely.  Each player must
choose a characteristic G or B and output a value in $\{0, 1,
\pass\}$.
Let $M > 6 + 6\epsilon$.
We define the utility function $u$ as follows.
\begin{itemize}
\item  If player 1 (the broadcaster) has type $v$, then
\begin{itemize}
\item if there exists a set $S$ of at least $n-t$ players such that
all players in $S$ choose G and output $v$, and the broadcaster
chooses G, then the broadcaster gets 1;
\item if there does not exist a set $R$ of at least $n-t$ players
and a value $v' \in \{0,1\}$ such that all players in $R$ choose
characteristic G and output a value in $\{v', \pass\}$, and the
broadcaster chooses characteristic B, then the broadcaster gets $M$;

\item in all other cases, the broadcaster gets 0.

\end{itemize}

\item Utility for player $i \ne 1$:
\begin{itemize}
\item if there exists a set $R$ of at
least $n-(k+t)$ players and a value $v' \in \{0,1\}$ such that all
players in $R$ choose G and all players in $R$ output a value in
$\{v', \pass\}$, then
player $i$ gets $1$ if he chooses G and
outputs a value in $\{v', \pass\}$ and gets 0 otherwise;

\item in all other cases, if $i$ chooses $B$ he gets $M$ and if he
chooses G he gets 0.
\end{itemize}
\end{itemize}

Consider a mediator that receives a value from the broadcaster and
sends this value to all players. It is easy to see that the strategy
$\sigma$ where the broadcaster truthfully tells the mediator his
type and chooses characteristic G, and all other players choose
characteristic G and output the value sent by the mediator, is a
$(k,t)$-robust equilibrium whose payoff is 1 for all players.

We claim that we cannot $\epsilon$-implement this mediator using
cheap talk if $n \le k+3t$. Suppose, by way of contradiction, that
there exists a strategy $\sigma'$ in $\Gammact$ that
$\epsilon$-implements $\sigma$. Much as in the proof of
\theoremref{thm:lowerbound-2}, we can transform $\sigma'$ into a
protocol $P$ for the $(k,t)$-partial broadcast problem: take $P_i$
to be the strategy where process $i$ sends messages according to
$\sigma'_i$, taking its initial value to be its type if $i=1$, and
decides on the value output by $\sigma'_i$.

Viewing $P$ as a protocol for the $(k,t)$-partial broadcast problem,
let $T$ be the set of processes guaranteed to exist by
\propref{prop:Byzgen}. If $|T|\leq t$, $1 \notin T$ and the
execution $(P_{N-T}, \Psi(P)_T)$ fails the agreement condition with
probability at least 1/3, then it is easy to see that $\sigma'$ is
not $\epsilon$--$t$-immune.
Otherwise, if $|T|\leq k+t$, $1\in T$ and the execution $(P_{N-T},
\Psi(P)_T)$ fails the no-disagreement condition with probability at
least 1/6, then it is easy to see that in $\Gamma$, deviating to
$\Psi(P)$ and choosing  B gives the members of $T$ an expected
utility greater than $M/6=1+\epsilon$, contradicting the assumption
the $\sigma'$ is $\epsilon-(k,t)$-robust.
\end{proof}

\subsection{Proof of
\theoremref{thm:d-lowerbound}}\label{sec:proof-d-lowerbound}

\rethm{thm:d-lowerbound}
If $k + t < n \le 2(k + t) $
and $k \ge 1$,
then there exists
a game  $\Gamma$, a mediator game $\Gamma_d$ that extends $\Gamma$,
a strategy $\sigma$ in $\Gamma_d$, and a strategy $\rho$ in $\Gamma$
such that
\begin{itemize}
\item[(a)] for all $\epsilon$ and $b$, there exists a
utility function $u^{\,b,\epsilon}$  such that $\sigma$ is a
$(k,t)$ robust equilibrium in
$\Gamma_d(u^{\,b,\epsilon})$ for all $b$ and $\epsilon$, $\rho$ is a
$(k,t)$-punishment strategy with respect to  $\sigma$ in
$\Gamma(u^{\,b,\epsilon})$ if $n > k + 2t$,  and there does not
exist an \eps $(k,t)$-robust implementation of $\sigma$ that runs in
expected time $b$ in the cheap-talk extension
$\Gammact(u^{b, \epsilon})$ of $\Gamma(u^{b, \epsilon})$,
\item[(b)] there exists a utility function
$u$ such that  $\sigma$ is a
$(k,t)$ robust equilibrium in
$\Gamma_d(u)$ and, for all $b$, there exists $\epsilon$ such that
there does not exist an \eps $(k,t)$-robust implementation of
$\sigma^i$ that runs in expected time $b$ in the cheap-talk
extension $\Gammact(u)$ of $\Gamma(u)$.
\end{itemize}
This is true even if players are computationally bounded, we assume
cryptography and there are broadcast channels. \erethm

\begin{proof}

First assume that $k=1$, $t=0$, and $n=2$. Consider a 2-person
secret-sharing game $\Gamma$ with the secret taken from the field $F
= \{0, \ldots, 6\}$, and the shares are signed using check vectors.
Specifically, nature uniformly chooses a secret $s \in F$ and a
degree $2$ polynomial $f$ over $F$ such that $f(0)=s$ and all
remaining coefficients are uniformly random. Nature also chooses for
$i \in \{1,2\}$, check vectors: $y_i$ uniformly in $F$, $b_i$
uniformly in $F \setminus \{0\}$ and $c_i$ such that $f(i)+y_i
b_i=c_i$.
Let $\bar{i} = 3-i$;
thus, $\bar{i}$ is the player other than $i$.

Player $i \in \{1,2\}$ gets as input $(f(i), y_i, b_{\bar{i}},
c_{\bar{i}})$ and must guess the secret. Let $u^{M,\delta}_i$ be the
following utility function for player $i \in \{1,2\}$: If player $i$
gets the right answer (i.e., guesses the secret) and $\bar{i}$ does
not, then $i$ gets $M$; if both get the right answer, then $i$ gets
1; if $\bar{i}$ gets the right answer and $i$ does not, then $i$
gets $-M + 2 -2\delta$; finally, if neither get the right answer,
then $i$ gets $1-\delta$.

Consider the following mediator. It expects to receive from each
player $i$ 4 field values: a share $a_i$, a signature $y_i$, and two
verification values $b_{\bar{i}}, c_{\bar{i}}$. If any player does
not send 4 values then the mediator chooses a value in $F$ at
random, and sends that to both players. Otherwise the mediator
interpolates the degree 1 polynomial $f$ from the shares
$a_1$ and $a_2$. Then it checks that $a_i+y_i b_i=c_i$ for $i \in \{1,2\}$.
If both checks are successful, he sends $f(0)$ to both players,
otherwise he sends both a value in $F \setminus \{f(0)\}$ chosen
uniformly at random.

Consider the truthful strategy $\sigma$ in the mediator game:
players tell the mediator the truth, and the mediator reports the
secret. The strategy profile $\sigma$ gives both players utility 1 for
all utility function $u^{M,\delta}$.  Truthfulness is easily seen to
be a 1-robust strategy in the mediator game (i.e., a Nash
equilibrium).  If a player $i$ lies and $\bar{i}$ tells the truth,
then with probability $6/7$, $i$ will be caught.
In this case, $i$ will definitely play the wrong value.
(Note that $i$ will play the right value if $f(0)$ is sent,
since $i$ will
be able to calculate what the true secret should have been, given his
lie; his calculation will be incorrect if a value other than $f(0)$ is
sent, which is what happens if $i$'s lie is detected.)
On the other hand, $\bar{i}$ will play the right value with probability
$1/6$.
Thus, $i$'s expected utility if he is caught is
$5(1-\delta)/6 + (-M + 2 -2\delta)/6 = (7-7\delta - M)/6$.  If
player $i$ is not caught, then his utility is
$M$.
Thus, cheating
has expected utility
$1-\delta$, so $i$ does not gain by lying as
long as $\delta \ge 0$.

Moreover, it is easy to see that, if $\delta > 0$ and $i$ chooses a
value at random, then if player $\bar{i}$ chooses the same value,
his expected utility is $1/7 + 6/7(1-\delta) = 1 - 6\delta/7$; if
player $\bar{i}$ chooses a different value, then his expected
utility is $M/7 + (-M + 2 - 2\delta)/7 + (1-\delta)(5/7) =
1-\delta$; it follows that player $\bar{i}$'s expected utility is at
most $1-6\delta/7$. Thus, if $\delta > 0$, then choosing a value at
random is a 1-punishment strategy with respect to $\sigma$ in
$\Gamma(u_{M,\delta})$ (even if the other player knows what value is
chosen).

For part (a), fix $\epsilon > 0$ and $b$.  We show that there is no
cheap-talk strategy $\sigmact $ that $\epsilon$-implements
$\sigma$ and has expected running time $b$ in
$\Gammact(u^{M,\epsilon})$ if $M > (1-\epsilon) + 28b\epsilon/3$.
Suppose, by way of contradiction, that there exists such a
cheap-talk strategy $\sigmact $. The key idea is to consider the
expected probability that a player will be able to guess the correct
answer at any round, assuming that both players use $\sigmact $.
With no information, the probability that $i$ guesses the right
answer is $1/7$
and all
values
are equally likely. When the strategy terminates, the probability
must be 1 (because we assume that both players will know the right
answer at the end if
they follow
the recommended strategy). In
general, at round $j$, player $i$ has acquired some information
$I^j$. (What $I^j$ is may depend on the outcome of coin tosses, of
course.) There is a well-defined probability of guessing the right
answer given $I^j$.  Thus, the expected probability of player $i$
guessing the right answer after round $j$ is the sum, taken over all
the possible pieces of information $I^j$ that $i$ could have at the
end of round $j$, of the probability of getting information $I^j$
times the probability of guessing the right answer given $I^j$.  By
Markov's inequality, both players terminate by round $2b$ with
probability at least $1/2$, the expected probability of guessing the
right answer by round $2b$ must be at least $4/7$ for both players.
(Since if an execution terminates, he can guess the right answer
with probability 1; otherwise, he can guess it with probability at
least $1/7$.)   Thus, for each player $i$, there must be a round $b'
< b$ such that the expected probability of player $i$ getting the
right answer increases by at least $3/7b$ between round $b'$ and
$b'+1$.  It follows that there must be some round $b' < b$ such that
either the expected probability of
player
1 guessing the answer after round
$b'+1$ is at least $3/14b$ more than that of
player
2 guessing the answer
at round after round $b'$, or the expected probability
player
2 guessing
the answer after round $b'+1$ is at least $3/32b$ more than that of
player
1 guess the answer after round $b'$.   (Proof: Consider a round $b'$
such that player 1's expected probability of guessing the right
answer increases by at least $3/7b$.  If player 2's probability of
guessing the right answer is at least $3/14b$ more than that of
player 1 at round $b'$, then clearly player 2's probability of
guessing the right answer at round $b'$ is at least $3/14b$ more
than player 1's probability of guessing the right answer at $b'-1$;
otherwise, player 1's probability of guessing the right answer at
round $b'+1$ is at least $3/14b$ more than player 2's probability of
guessing it at round $b'$.)

Suppose, without loss of generality, that $b'$ is such that player
1's probability of guessing the right answer at round $b'+1$ is at
least $3/14b$ more than player 2's probability of guessing the right
answer at round $b'$.  Then player 1 deviates from $\sigmact $ by
not sending any messages to player 2 at round $b'$, and then making
a decision based on his information at round $b'+1$, using his best
guess based on his information.  (Note that player 2 will still send
player 1 a message at round $b'$ according to $\sigmact $.) The
best player 2 can do is to use his round $b'$ information.  If
$\alpha_1$, $\alpha_2$, $\alpha_3$, and $\alpha_4$ are the
probabilities of
player
1 getting the right answer and
player
2 not, both getting
the right answer, 2 getting the right answer and 1 not, and neither
getting the right answer, we must have $\alpha_1 - \alpha_3 \ge
3/14b$.  Moreover,
since $\vecsigmact$ is an $\epsilon$-implementation of $\sigma$, by
assumption, the expected utility of player $1$
if he deviates is at least $(1-\epsilon)(1 - (\alpha_1 - \alpha_3))
+ M(\alpha_1 - \alpha_3)$. It easily follows that, since $M >
(1-\epsilon) + 28b\epsilon/3$, then 1's expected utility by
deviating at round $b'$ is greater than $1+\epsilon$.  Hence,
$\sigma$ is not an $\epsilon$-equilibrium in
$\Gammact(u^{M,\epsilon})$.

For part (b), consider the utility function $u^{2,0}$. Note that
there is no 1-punishment strategy $\Gammact(u^{2,0})$ with
respect to $\sigma$.  We show that, for all $b$, there is no
cheap-talk strategy $\sigmact $ that $\epsilon$-implements
$\sigma$ and has expected running time $b$ in $\Gammact(u^{2,0})$
if $\epsilon < 3/14b$.  Suppose that $\sigmact $ is a cheap-talk
that $\epsilon$-implements $\sigma$ and has expected running time
$b$.  The argument above shows that there must be a round $b'$ where
one of player 1 or player 2 can deviate and have expected utility at
least $(1 - (\alpha_1 - \alpha_3)) + 2(\alpha_1 - \alpha_3) = 1 +
(\alpha_1 - \alpha_3) > 1 +  3/14b$.  The result immediately
follows.

For the general argument, we do the proof of part (a) here; the
modifications needed to deal with part (b) are straightforward.
Consider a $k+t+1$ out of $n$ secret sharing game, where the initial
shares are ``signed'' using check vectors. Specifically, for each
share $f(i)$ and for each player $j \in N \setminus {i}$, player $i$
is given a uniformly random value $y_{ij}$ in $F$ and player $j$ is
given a uniformly random value $b_{ij}$ in $F \setminus \{0\}$ and a
value $c_{ij}$ such that $f(i)+y_{ij}b_{ij}=c_{ij}$. In the
underlying game, players can either choose a value in the field
(intuitively, their best guess as to the secret) or play $\detect$.
The utility functions are defined as follows:
\begin{itemize}
\item if at least $n-t$ players play $\detect$, then all players playing
$\detect$ get 1, and all others get 0;

\item if fewer than $n-t$ players play $\detect$ and at least $n-(k+t)$
but fewer than $n-t$ players play the secret, then the players playing
the secret get $M$ and the other players get $-M+2 - 2\delta$

\item if fewer than $n-t$ players play $\detect$ and either $n-t$ or
more players or fewer than $n-(k+t)$ players play the secret, then
the players playing the secret get 1, and the remaining players get
$1-\delta$.
\end{itemize}

In the mediator game, each player is supposed to send the mediator
his type (share, signatures, and verifications).
The mediator checks that all the shares sent pass the  checks. Note
that each share can be subjected to $n$ checks, one for each player.
A share is \emph{reliable} if it passes at least $n-t$ checks.  If
there are not at least $n-t$ reliable shares, but a unique
polynomial $f$ can be interpolated through the shares that are sent,
then the mediator sends a random value that is not $f(0)$ to all the
players; otherwise, the mediator chooses a value in $F$ at random
and sends it to all the players. If there are at least $n-t$
reliable shares, the mediator checks if a unique polynomial of
degree $k+t$ can be interpolated through the shares.  If so, the
mediator sends the secret to all the players; if not, the mediator
sends $\detect$ to all the players.  Let $\sigma$ be the strategy
for the players where they truthfully tell the mediator their type,
and play what the mediator sends.

We claim that playing 1 is a $(k+t)$-punishment strategy with
respect to $\sigma$ if $\delta > 0$ and $n> k + 2t$, and that
$\sigma$ is a $(k,t)$-robust equilibrium. The argument that playing
1 is a $(k+t)$-punishment strategy is essentially identical to the
argument for the 2-player game. However, note that if $n \le k  +
2t$, then $t \ge n - (k+t)$.  If at most $t$ players play 1, this is
not a punishment strategy since the remaining $n-t$ players can play
$\detect$ and guarantee themselves a payoff of 1.

To show that $\sigma$ is a $(k,t)$-robust equilibrium, we first show
it is $t$-immune. Suppose that a subset $T$ of at most $t$ players
attempt to fool the mediator by guessing shares and appropriate
check values, and the remaining players play $\sigma$.  Then at
least $n-t$ shares will be reliable.  Note that $n-t \ge k+t+1$.
Either the mediator can interpolate a unique polynomial $f$ of
degree $k+t$ through the reliable shares or not.  In the former
case, the good players will learn the secret; in the latter case,
the good players will play $\detect$.  In either case, their payoff
is 1.  Thus, $\sigma$ is $t$-immune.

For robustness, suppose that a set $T$ up to $k+t$ players deviate
from the recommended strategy.   If the true polynomial is $f$, the
players in $T$ can convince the mediator that some polynomial $f'
\ne f$ is the true polynomial, and the players in $T$ know $(k+t)$
of the points on $f'$, then, once the players in $T$ learn $f'(0)$,
they will know $k+t+1$ points on $f'$, and hence will be able to
compute $f'$. They can then compute the shares of all the other
players, and thus compute $f(0)$.  This can happen only if $|T| =
k+t$, $2(k+t) - n$ of the players in $T$ send their true shares (and
the correct check vectors), and the remaining players in $T$ send
incorrect shares. So we assume that $|T| = k+t$, and $n-(k+t)$
players in $T$ send incorrect values.  If the mediator cannot
interpolate a unique polynomial through the values sent, then the
mediator chooses a value in $F$ at random and sends it to all the
players.  Even if the players in $T$ know that this is what
happened, the players not in $T$ are playing a punishment strategy,
so a player in $T$ cannot get expected utility higher than $1 -
6\delta/7$, even if they know that the mediator is sending a random
value. If the mediator can interpolate a unique polynomial $f$
through the shares sent, then the mediator will send $f(0)$ if all
of the $n-(k+t)$ shares received are reliable, and a value different
from $f(0)$ otherwise.  In the former case, which occurs with
probability $1/7^{n-(k+t)}$, the players in $T$ can compute the true
secret, and will get a payoff of $M$.  In the latter case, they
compute the wrong value.  With probability $(1/6)(1-1/7^{n-(k+t)})$,
the other players get the right value and the players in $T$ get a
payoff of $-M+2 - 2\delta$; otherwise, they get a payoff of 1.  It
is easy to see that the expected utility of the players in $T$ is at
most $1-2\delta/7$, so the rational players will not deviate.

Suppose that this mediator strategy can be implemented using cheap
talk.  We claim that, as in the proof of
\theoremref{pro:lowerboundbii}, we can use the implementation to give
a cheap-talk implementation in the 2-player game.  We simply divide the
players into three groups: groups $A$ and $B$ both have $n - (k+t)$
members; group $K$ has the remaining $2(k+t) - n$ players (group $K$ may
be empty).  Notice that $|A \cup B| = |A \cup K| = k+t$.
Just as in the 2-player case, we can show that there must be
a round $b'$ such that if the players in group $A$ pool their
information together, the probability of them guessing the right answer
at round $b'+1$
is at least $3/14b$ more than the probability of the players in group
$B$ guessing the answer after round $b'$ even if the players in group
$B$ pool their knowledge together, or the same situation holds with the
roles of $A$ and $B$ reversed.  Assume that $A$ is the group that has
the higher probability of guessing the right answer at time $b'$.  Then
we assume that the players in $A \cup K$ deviate
by not sending a round $b'$ message to the players in $B$.  (Here we use
the fact that $|A \cup B| \le k+t$.)  Now the argument continues as in
the 2-player case.
\end{proof}

\newpage
\bibliographystyle{alpha}
\bibliography{game1}

\begin{thebibliography}{FHHW03}

\bibitem[ADGH06]{ADGH06}
I.~Abraham, D.~Dolev, R.~Gonen, and J.~Y. Halpern.
\newblock Distributed computing meets game theory: Robust mechanisms for
  rational secret sharing and multiparty computation.
\newblock In {\em Proc.~25th ACM Symp.~Principles of Distributed Computing},
  pages 53--62, 2006.

\bibitem[ADGH07]{ADGH06full}
I.~Abraham, D.~Dolev, R.~Gonen, and J.~Y. Halpern.
\newblock Distributed computing meets game theory: Robust mechanisms for
  rational secret sharing and multiparty computation.
\newblock unpublished manuscript, 2007.

\bibitem[ADH]{ADH07a}
I.~Abraham, D.~Dolev, and J.Y. Halpern.
\newblock On implementing mediators with asynchronous cheap talk.
\newblock Unpublished manuscript.

\bibitem[AH03]{AH03}
R.~J. Aumann and S.~Hart.
\newblock Long cheap talk.
\newblock {\em Econometrica}, 71(6):1619--1660, 2003.

\bibitem[Aum59]{Aumann59}
R.~J. Aumann.
\newblock Acceptable points in general cooperative $n$-person games.
\newblock {\em Contributions to the Theory of Games, Annals of Mathematical
  Studies}, IV:287--324, 1959.

\bibitem[Aum87]{Aumann87}
R.~J. Aumann.
\newblock Correlated equilibrium as an expression of {B}ayesian rationality.
\newblock {\em Econometrica}, 55:1--18, 1987.

\bibitem[Bar92]{Barany92}
I.~Barany.
\newblock Fair distribution protocols or how the players replace fortune.
\newblock {\em Mathematics of Operations Research}, 17:327--340, 1992.

\bibitem[Ben03]{Bp03}
E.~Ben{-}Porath.
\newblock Cheap talk in games with incomplete information.
\newblock {\em J.~Economic Theory}, 108(1):45--71, 2003.

\bibitem[BGW88]{BGW88}
M.~Ben{-}Or, S.~Goldwasser, and A.~Wigderson.
\newblock Completeness theorems for non-cryptographic fault-tolerant
  distributed computation.
\newblock In {\em Proc.~20th ACM Symp.~Theory of Computing}, pages 1--10, 1988.

\bibitem[BN00]{BN00}
Dan Boneh and Moni Naor.
\newblock Timed commitments.
\newblock In {\em CRYPTO '00: Proceedings of the 20th Annual International
  Cryptology Conference on Advances in Cryptology}, pages 236--254.
  Springer-Verlag, 2000.

\bibitem[BPW89]{BernheimPelegWhinston}
B.~D. Bernheim, B.~Peleg, and M.~Whinston.
\newblock Coalition proof {N}ash equilibrium: Concepts.
\newblock {\em J.~Economic Theory}, 42(1):1--12, 1989.

\bibitem[CCD88]{CCD88}
D.~Chaum, Claude Cr{\'e}peau, and I.~Damgard.
\newblock Multiparty unconditionally secure protocols.
\newblock In {\em Proc.~20th ACM Symp.~Theory of Computing}, pages 11--19,
  1988.

\bibitem[CS82]{CS82}
V.~P. Crawford and J.~Sobel.
\newblock Strategic information transmission.
\newblock {\em Econometrica}, 50(6):1431--51, 1982.

\bibitem[DHR00]{DHR00}
Y.~Dodis, S.~Halevi, and T.~Rabin.
\newblock A cryptographic solution to a game theoretic problem.
\newblock In {\em CRYPTO 2000: 20th International Cryptology Conference}, pages
  112--130. Springer-Verlag, 2000.

\bibitem[EGL85]{EGL85}
S.~Even, O.~Goldreich, and A.~Lempel.
\newblock A randomized protocol for signing contracts.
\newblock {\em Commun. ACM}, 28(6):637--647, 1985.

\bibitem[Eli02]{Eliaz00}
K.~Eliaz.
\newblock Fault-tolerant implementation.
\newblock {\em Review of Economic Studies}, 69(3):589--610, 2002.

\bibitem[FHHW03]{FHHW03}
M.~Fitzi, M.~Hirt, T.~Holenstein, and J.~Wullschleger.
\newblock Two-threshold broadcast and detectable multi-party computation.
\newblock In {\em Advances in Cryptology --- EUROCRYPT '03}, volume 2656 of
  {\em Lecture Notes in Computer Science}, pages 51--67. Springer-Verlag, 2003.

\bibitem[FLM86]{FLM86}
M.~J. Fischer, N.~A. Lynch, and M.~Merritt.
\newblock Easy impossibility proofs for distributed consensus problems.
\newblock {\em Distributed Computing}, 1(1):26--39, 1986.

\bibitem[FLP85]{FLP}
M.~J. Fischer, N.~A. Lynch, and M.~S. Paterson.
\newblock Impossibility of distributed consensus with one faulty processor.
\newblock {\em J. ACM}, 32(2):374--382, 1985.

\bibitem[For90]{F90}
Francoise Forges.
\newblock Universal mechanisms.
\newblock {\em Econometrica}, 58(6):1341--64, 1990.

\bibitem[GK06]{GK06}
D.~Gordon and J.~Katz.
\newblock Rational secret sharing, revisited.
\newblock In {\em SCN (Security in Communication Networks) 2006}, pages
  229--241, 2006.

\bibitem[GMW87]{GMW87}
O.~Goldreich, S.~Micali, and A.~Wigderson.
\newblock How to play any mental game.
\newblock In {\em Proc.~19th ACM Symp.~Theory of Computing}, pages 218--229,
  1987.

\bibitem[Gol04]{goldreich03}
O.~Goldreich.
\newblock {\em Foundations of {C}ryptography, {V}ol. 2}.
\newblock Cambridge University Press, 2004.

\bibitem[Hel05]{Heller05}
Y.~Heller.
\newblock A minority-proof cheap-talk protocol.
\newblock Unpublished manuscript, 2005.

\bibitem[HT04]{HT04}
J.~Y. Halpern and V.~Teague.
\newblock Rational secret sharing and multiparty computation: extended
  abstract.
\newblock In {\em Proc.~36th ACM Symp.~Theory of Computing}, pages 623--632,
  2004.

\bibitem[IML05]{IML05}
S.~Izmalkov, S.~Micali, and M.~Lepinski.
\newblock Rational secure computation and ideal mechanism design.
\newblock In {\em Proc.~46th IEEE Symp.~Foundations of Computer Science}, pages
  585--595, 2005.

\bibitem[KW82]{KW82}
D.~M. Kreps and R.~B. Wilson.
\newblock Sequential equilibria.
\newblock {\em Econometrica}, 50:863--894, 1982.

\bibitem[KY84]{KY84}
A.~Karlin and A.~C. Yao.
\newblock Probabilistic lower bounds for byzantine agreement.
\newblock Unpublished manuscript, 1984.

\bibitem[Lam83]{L83}
L.~Lamport.
\newblock The weak byzantine generals problem.
\newblock {\em J. ACM}, 30(3):668--676, 1983.

\bibitem[LMPS04]{LMPS04}
M.~Lepinski, S.~Micali, C.~Peikert, and A.~Shelat.
\newblock Completely fair {SFE} and coalition-safe cheap talk.
\newblock In {\em Proc.~23rd ACM Symp.~Principles of Distributed Computing},
  pages 1--10, 2004.

\bibitem[LMS05]{LMS05}
M.~Lepinksi, S.~Micali, and A.~Shelat.
\newblock Collusion-free protocols.
\newblock In {\em Proc.~37th ACM Symp.~Theory of Computing}, pages 543--552,
  2005.

\bibitem[LSP82]{LSP82}
L.~Lamport, R.~Shostak, and M.~Pease.
\newblock The {B}yzantine {G}enerals problem.
\newblock {\em ACM Trans.~on Programming Languages and Systems}, 4(3):382--401,
  1982.

\bibitem[LT06]{LT06}
A.~Lysyanskaya and N.~Triandopoulos.
\newblock Rationality and adveresarial behavior in multi-party comptuation.
\newblock In {\em CRYPTO 2006}, pages 180--197, 2006.

\bibitem[MW96]{MW96}
D.~Moreno and J.~Wooders.
\newblock Coalition-proof equilibrium.
\newblock {\em Games and Economic Behavior}, 17(1):80--112, 1996.

\bibitem[Mye97]{M97}
Roger~B. Myerson.
\newblock {\em Game Theory: Analysis of Conflict}.
\newblock Harvard University Press, September 1997.

\bibitem[PW96]{PW96}
B.~Pfitzmann and M.~Waidner.
\newblock Information-theoretic pseudosignatures and byzantine agreement for $t
  >= n/3$.
\newblock Technical Report RZ 2882 (\#90830), IBM Zurich Research Laboratory,
  1996.

\bibitem[Rab]{R81}
M.~Rabin.
\newblock How to exchange secrets with oblivious transfer.
\newblock 1981. http://eprint.iacr.org/2005/187.

\bibitem[RB89]{RB89}
T.~Rabin and M.~Ben{-}Or.
\newblock Verifiable secret sharing and multiparty protocols with honest
  majority.
\newblock In {\em Proc.~21st ACM Symp.~Theory of Computing}, pages 73--85,
  1989.

\bibitem[Sel75]{Selten75}
R.~Selten.
\newblock Reexamination of the perfectness concept for equilibrium points in
  extensive games.
\newblock {\em International Journal of Game Theory}, 4:25--55, 1975.

\bibitem[SRA81]{SRA81}
A.~Shamir, R.~L. Rivest, and L.~Adelman.
\newblock Mental poker.
\newblock In D.~A. Klarner, editor, {\em The Mathematical Gardner}, pages
  37--43. Prindle, Weber, and Schmidt, Boston, Mass., 1981.

\bibitem[UV02]{UV02}
A.~Urbano and J.~E. Vila.
\newblock Computational complexity and communication: Coordination in
  two-player games.
\newblock {\em Econometrica}, 70(5):1893--1927, 2002.

\bibitem[UV04]{UV04}
A.~Urbano and J.~E. Vila.
\newblock Computationally restricted unmediated talk under incomplete
  information.
\newblock {\em Economic Theory}, 23(2):283--320, 2004.

\bibitem[Yao82]{yao:sc}
A.~Yao.
\newblock Protocols for secure computation (extended abstract).
\newblock In {\em Proc.~23rd IEEE Symp.~Foundations of Computer Science}, pages
  160--164, 1982.

\end{thebibliography}

\end{document}